%% file: paper_multiscale_bloodflow.tex
\documentclass{amsart}

\usepackage[super,comma,sort&compress]{natbib}
\makeatletter
\newcommand*{\citenst}[2][]{%
	\begingroup
	\let\NAT@mbox=\mbox
	\let\@cite\NAT@citenum
	\let\NAT@space\NAT@spacechar
	\let\NAT@super@kern\relax
	\renewcommand\NAT@open{[}%
	\renewcommand\NAT@close{]}%
	\citet[#1]{#2}%
	\endgroup
}
\makeatother

\usepackage{mathtools,bm,comment,amssymb,enumitem,hyperref,xcolor,array,float,subcaption,float,pgfplots,etoolbox}
\usepackage[capitalise,nameinlink]{cleveref}
\usepackage[english]{babel}
\usepackage[margin=1in]{geometry}
\usepackage[foot]{amsaddr}
\usepackage[linesnumbered,ruled]{algorithm2e}
\usepackage[noend]{algpseudocode}
\usepackage[locale=DE]{siunitx}
\usepackage{textcomp}
\usepackage{placeins}
\usetikzlibrary{shapes,arrows}
\usepackage{todonotes}
\usepackage{units}
\usepackage{todonotes}
\usepackage{caption}
\usepackage{subcaption}
\usepackage{enumitem}
\usepackage{lipsum}
\usepackage{amsfonts}
\usepackage{graphicx}
\usepackage{epstopdf}
\usepackage{caption}
\usepackage{subcaption}
\usepackage{enumitem}
\usepackage{amsopn}

\usepackage{units}
\let\originalleft\left
\let\originalright\right
\renewcommand{\d}{\textup{d}}
\renewcommand{\rho}{\varrho}
\renewcommand{\left}{\mathopen{}\mathclose\bgroup\originalleft}
\renewcommand{\right}{\aftergroup\egroup\originalright}

\makeatletter
\@namedef{subjclassname@2020}{\textup{2020} Mathematics Subject Classification}
\makeatother

\title[Simulating blood flow and transport in breast tissue]{A 1D-0D-3D coupled model for simulating blood flow \\ and transport processes in breast tissue}

\author{Marvin Fritz$^1$, Tobias K\"oppl$^{1}$, J. Tinsley Oden$^2$, Andreas Wagner$^{1,*}$, \\ Barbara Wohlmuth$^{1,3}$, Chengyue Wu$^{2}$}
\thanks{${}^*$Corresponding author}
\email{\{fritzm, koepplto, wagneran, wohlmuth\}@ma.tum.de}
\email{oden@oden.utexas.edu}
\email{cw35926@utexas.edu}

\begin{document}

\maketitle
\vspace*{-2mm}
\begin{center} \footnotesize
	$^1$Department of Mathematics, Technical University of Munich, Germany \\
	$^2$Oden Institute for Computational Engineering and Sciences, The University of Texas at Austin, USA 
\end{center}

\begin{abstract}
In this work, we present mixed dimensional models for simulating blood flow and transport
processes in breast tissue and the vascular tree supplying it. These processes are considered, to start from the aortic inlet to the capillaries and tissue of the breast. Large variations in biophysical properties and flow conditions exist in this system necessitating the use of different flow models for different geometries and flow regimes. Large variations in biophysical properties and flow conditions exist in this system necessitating the use of different flow models for different geometries and flow regimes. In total, we consider four different model types. First, a system of 1D nonlinear hyperbolic PDEs is considered to simulate blood flow in larger arteries with highly elastic vessel walls. Second, we assign 1D linearized hyperbolic PDEs to model the smaller arteries with stiffer vessel walls. The third model type consists of ODE systems (0D models). It is used to model the arterioles and peripheral circulation. Finally, homogenized 3D porous media models are considered to simulate flow and transport in capillaries and tissue within the breast volume. Sink terms are used to account for the influence of the venous and lymphatic systems. Combining the four model types, we obtain two different 1D-0D-3D coupled models for simulating blood flow and transport processes: The first 1D-0D-3D model covers the whole path from the aorta to the breast, while the second model is a sub-model obtained by restriction to breast vasculature and tissue making possible a significant reduction in computational cost. Several numerical experiments are conducted that demonstrate realistic flow simulations compared to existing data on blood flow in human breast and vascular system.
\end{abstract}

\input{multiscale_1_introduction.tex}

\input{multiscale_2_model_setting.tex}

\input{multiscale_3_model_components.tex}

\input{multiscale_4_coupling_and_boundary_conditions.tex}

\input{multiscale_5_coupled_models.tex}

\input{multiscale_6_numerical_solution_methods.tex}

\input{multiscale_7_results.tex}

\input{multiscale_8_conclusions.tex}

\section*{Acknowledgements}
The authors gratefully acknowledge the support of DFG (IGSSE, WO 671/11-1, WO 671/20-1) and U.S. Department of Energy, Office of Science, Office of Advanced Scientific Computing Research, Mathematical Multifaceted Integrated Capability Centers (Award DE-SC0019303).

{\small \bibliography{literature}
\bibliographystyle{abbrv}}

\end{document}

%% file: multiscale_1_introduction.tex
\section{Introduction}

The development of computational models for the simulation of blood flow and transport processes within the human cardiovascular system has become an important goal in modern computational medicine. Accurate predictions of flow and transport processes can provide non-invasive options to study and design effective medical procedures for treating several diseases including breast cancer.\cite{d2007multiscale,dewhirst2017transport} Invasive medical procedures such as biopsy not only cause damage to healthy tissue but also may yield useful data only at one point in time. In silico modeling, on the other hand, is a valuable tool for studying the individual patient's pathophysiology throughout, systematically evaluating and forecasting the outcomes of candidate treatments, and, most importantly, personalizing healthcare. In addition, numerical simulation techniques are used for testing hypotheses, e.g., see References~\citenum{drzisga2016numerical,el2018investigating,koeppl2018numerical,possenti2019computational}.

A typical example of a prevalent therapeutic procedure in which a transport process plays a central role is systemic therapy for breast cancer patients. A port is placed in the upper vena cava during such procedures. Chemo-therapeutic drugs are injected through the port and delivered to the vasculature in the affected breast via the heart's, pulmonary circulation, and engage the larger arteries of the systemic circulation.\cite{butros2014direct,puel1993superior} It is critical to know which portion of the injected drugs reaches the tumor in order to estimate the efficacy of this therapy. With this information, the duration rate and dose of an infusion can be optimized so that both the tumor and any negative impact of the injected drug on healthy tissue are minimized.

To achieve a representation of systemic drug delivery with a high fidelity, it is necessary to create a model for blood flow and transport of a solute in the heart chambers, pulmonary circulation, arteries branching out of the heart, as well as the arterial vessels in the breast. In addition, flow and transport models for the vasculature have to be coupled with the corresponding models for the breast tissue that is supplied by the vasculature and contains the tumor. Because flow characteristics vary depending on the geometry and mechanical properties of various vascular components, developing a computer model of such complex processes is a challenging undertaking. We describe and implement such flow and transport models in this paper. 

Due to the complexity of the vascular tree, it is not feasible to model cardiovascular systems using conventional flow models. The diameters of the larger systemic arteries are centimeters, while capillary diameters are micrometers. Further, large systemic arteries experience turbulent flow while capillary flow has low Reynolds numbers and velocities. To capture the wide variations in both time and length scales in the cardiovascular systems, we employ when feasible simplified flow models of different dimensions and complexity.

The various flow behaviors and vessel features appearing in our vasculature are represented by a combination of 1D PDE systems, ODEs, and 3D flow models, which form a 1D-0D-3D coupled model.
A similar type of modeling approach has been applied, e.g., to model flow in aneurysms.\cite{ho2009hybrid,blanco2009potentialities}
To compute the flow field within an aneurysm and wall shear stresses at its surface, a three-dimensional (3D) flow model is used. 
Dimensional reduced models based on one-dimensional (1D) partial differential equations (PDEs) or ordinary differential equations (ODEs) are coupled with the 3D flow model to provide realistic boundary conditions. 
In contemporary literature, ODEs coupled with higher dimensional models are also referred to as ''0D models'', since they contain no space variable.  Mixed-dimensional models have also shown success in the simulation of flow and transport in vascularized tissue.\cite{koch2020modeling,kojic2017composite}
Here, the tissue matrix is considered as a 3D porous medium, whereas the blood vessel network supplying the tissue is approximated by a 1D graph, which results in a 3D-1D coupled flow model. 
This modeling approach has been used in Wu et al.\citenum{wu2020patient} to model flow and transport in a breast network consisting of smaller arteries as well as breast tissue. 

Flow and transport in the larger arteries branching out of the heart and the smaller arteries containing the breast are governed by 1D PDE systems. 
For the larger arteries we use a nonlinear hyperbolic equation to model pulsatile blood flow in vessels with elastic walls, see \cite{hughes1973one,vcanic2003mathematical}.
For the smaller arteries in the breast, a linearized version of this flow model is considered which is valid for small deformations of the vessel diameter. 
Preliminary research on the coupling of various 1D flow models appears in Drzisga et al.\cite{drzisga2016numerical} 
At the interface between the larger and smaller arteries, we must create new bidirectional coupling conditions. This coupling places a significant constraint on the time step size, which motivates a low-cost one-directional coupled model. Surrogate models that account for the influence of vessels that are not part of the 1D networks are required at the outlets of the 1D networks. We couple the corresponding 1D models with Windkessel models based on ODEs at the outlets of the nonlinear 1D network to incorporate the capacity and flow resistance of the vasculature located beyond the outlets.\cite{alastruey2008lumped,stergiopulos1996four} These models are commonly known as lumped-parameter models, see Fernandez et al.\cite{fernandez2005analysis} At the outlets of the linear network, we attach surrogate models for the arterioles that connect the breast network outlets to the capillary bed. 
New lumped parameter models are developed for these vessel trees. 

We consider 3D homogenized models such as those presented in References~\citenum{jozsa2021porous,padmos2021coupling,rohan2018modeling,shipley2020hybrid,peyrounette2018multiscale,vidotto2019hybrid} to determine the flow field within the capillaries and tissue. 
This results in a 3D-3D double continuum model of flow within the breast microvasculature and tissue. 
Both 3D models are based on the theory of porous media. 
Because the Reynolds numbers in both the capillaries and the tissue are less than one, Darcy's equation is used to calculate the pressures in both regions. 
We couple both equations by their source terms to model fluid exchange between the vascular system and the interstitial space of the tissue. 
By including sink terms in the Darcy equation for the capillary continuum, the influence of the venous system draining the breast tissue is accounted for. 
For tissue permeability, values from Wu et al.\cite{wu2020patient} are considered, whereas for capillary bed homogenization, existing techniques from the literature\cite{jonavsova2014complex,kremheller2021validation,kremheller2019approach,peyrounette2018multiscale,vidotto2019hybrid} are adopted. 

Finally, the model for arterioles is coupled with the capillary continuum. 
Unlike previous models,\cite{jonavsova2014complex,jozsa2021porous,padmos2021coupling,rohan2018modeling} we do not directly couple the 1D PDEs in the terminal vessels with the continuum, but we employ the lumped parameter model for the arterioles is coupled with both the 1D PDEs in the terminal vessels and the 3D continuum model for the capillaries. 
This is required because the pulsatile flow in the breast network must be converted into a non-pulsatile flow that is present in capillaries and tissue. To simulate the propagation of a solute, all flow models must be coupled to either a convection diffusion equation modeling solute concentrations.
This is done in References~\citenum{koppl2013reduced,masri2021reduced} for the larger arteries and the breast network. 
All other models, particularly the 0D or lumped parameter models, must be augmented by a transport process equation.

This paper is organized as follows: 
The generic and patient-specific blood vessel networks, as well as the modeling assumptions, are introduced in Section \ref{sec:modelSetting}.
Section \ref{sec:modelComponents} describes the mathematical sub-models that adhere to these assumptions and their coupling and boundary conditions are given in Section \ref{sec:coupling-and-boundary-conditions}. 
We combine this with our geometric data in Section \ref{sec:coupledModels}, resulting in
a complex 1D-0D-3D coupled model that contains nonlinear components. This model is referred to as a fully coupled model. Further, we also describe more efficient computational modeling approach that divides our data into a generic patient independent part that only needs to be calculated once and a patient specific part that needs to be recalculated for different breast geometries. Section \ref{sec:numsol} discusses our numerical solution methods, and Section \ref{sec:results} displays simulation results.  We compare the fully coupled model to available medical data and the simplified sub-model. Finally, closing remarks are given and an outlook for future work is described in Section \ref{sec:conclusion}.

%% file: multiscale_2_model_setting.tex
\section{Model setting}
\label{sec:modelSetting}

In this section, we describe our data sets and  formulate our most significant modeling assumptions.

\subsection{Description of the data sets}

\begin{figure} \centering
	\includegraphics[width=0.55\textwidth]{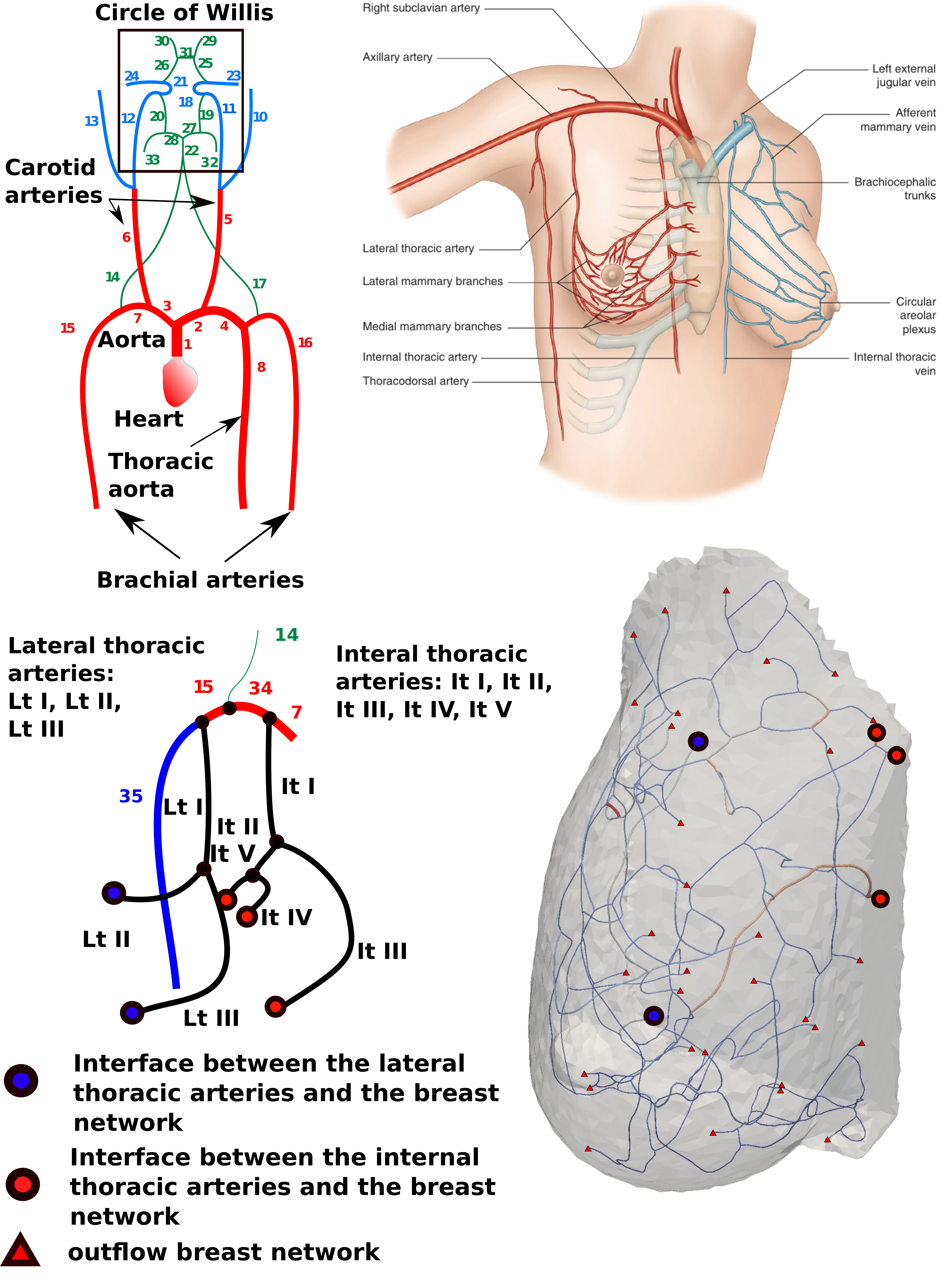}
	\caption{\label{fig:vascularTrees}Upper left: Idealized network consisting of the main arteries branching out of the heart and the Circle of Willis. Upper right: Vasculature supplying the breast tissue, reproduced from Fig.~13.4 from Barral et al.\cite{barral2011visceral} with permission from publisher. Lower left: Thoracic arteries branching out of the right subclavian (Vessel $7$) and brachial artery (Vessel $15$). As a result Vessel $7$\cite{alastruey2007modelling} is split into new vessels having the indices $7$ and $34$, while Vessel $15$ \cite{alastruey2007modelling} is split into $15$ and $35$. Lower right: A vascular network embedded inside a 3D breast volume. Connections between the thoracic arteries and the 1D breast network are blue for the lateral thoracic arteries and red for the interior thoracic arteries.}
\end{figure}

In order to simulate blood flow and transport processes in the larger systemic arteries branching out of the heart, we use the data\cite{alastruey2007modelling} as a generic patient unspecific geometry. These data describe an arterial network consisting of $33$ arteries and containing the aorta, carotid arteries as well as the Circle of Willis. Moreover, the main brachial arteries and the subclavian arteries are part of this network (see Figure \ref{fig:vascularTrees}, upper left). Table~$1$ in Alastruey et al.\cite{alastruey2007modelling} contains the average lengths, radii, wall thicknesses and elasticity parameters of the corresponding vessels. 
In particular, the subclavian arteries are of great interest for the objective of this paper since they contain several branches supplying blood and other constituents to the breast tissue (see Figure \ref{fig:vascularTrees}, upper right). 
Therefore, the right subclavian artery and brachial artery of the data set are split into two parts, which yields the network in the lower left corner of Figure~\ref{fig:vascularTrees} with the additional Vessels $34$ and $35$ and the additional branches (see Figure \ref{fig:vascularTrees}, lower left). Hereafter, this network (i.e., the network consisting of Vessel $1,\ldots,35$) is denoted as macrocirculation, and the set containing all its vessel ids is denoted by $I_{macro}$. The vessels with an outlet are contained in $I^{out}_{macro}$, while the only inlet at the aorta is assigned to $I^{in}_{macro}$.

As a patient specific geometry, we use the breast network that can be found in Wu et al.,\cite{wu2020patient,wu2019quantitative} see Figure \ref{fig:vascularTrees} (lower right).  First, a 3D volumetric mask of vasculature was segmented from high-resolution DCE-MRI.\cite{wu2020patient} A multiscale Hessian filter\cite{vignati2012fully} was used to subtract the pre-contrast from the post-contrast, high-spatial resolution DCE-MRI, calculating the probability of each voxel containing a vessel, and then a threshold was set to generate a binary mask of vasculature.\cite{wu2020patient} Second, the segmented vascular mask is skeletonized to its centerline, with the branching points, terminal ends, and vessels in the vascular network identified and the vessel orientation calculated for each vessel. After that, a gap filling process is used to transform the vascular network into a single connected graph.\cite{wu2019quantitative} Finally, a moving average of one imaging voxel size (i.e., 0.8 mm) is used to smooth out the vessel centerlines.

The breast network is considered as a graph-like structure, i.e., the data set comprises a list of 3D coordinates of the branching points, inlets and outlets. In the following, we will refer to them as 'network nodes'. Additionally, it contains information on the connectivity of the network nodes. 
By this means, the edges of the graph are defined. An average radius is assigned to each edge. 

Considering the size of the radii, it turns out that they range from $0.13\;\unit{mm}$ to $0.42\;\unit{mm}$. Thus it can be considered as a network of smaller arteries, see Table~1.1. in Formaggia et al.\cite{Quarteroni} In the following subsections, this network is denoted as the breast network, and its vessel indices are combined in a set $I_{breast}$. We choose the five largest arteries as inlets and collect their indices in $I^{in}_{breast}$ and the remaining terminal vessels in $I^{out}_{breast}$. 

There is still an information gap between the generic macrocirculation and the patient specific breast geometry which we close by introducing an extension network. According to standard medical textbooks, the blood vessels linking the subclavian arteries and the vasculature contained in breast tissue are called thoracic arteries, see Chapter~13 in Barral et al.\cite{barral2011visceral} and Figure \ref{fig:vascularTrees} (upper right) below. These consist of two main branches which are referred to as the internal and lateral thoracic arteries. Our construction of the thoracic network is depicted on the right of Figure~\ref{fig:vascularTrees} together with its connections to the macrocirculation and breast networks. We assign the indices of the vessels Lt I and It I to $I_{tho,1}$, Lt II and It II to $I_{tho,2}$ and the remaining vessels to $I_{tho,3}$ such that $I_{tho} = I_{tho,1} \cup I_{tho,2} \cup I_{tho,3}$ contains all indices of the extension or the thoracic vessels.

\subsection{Model assumptions}

\begin{figure} \centering
	\includegraphics[width=0.55\textwidth]{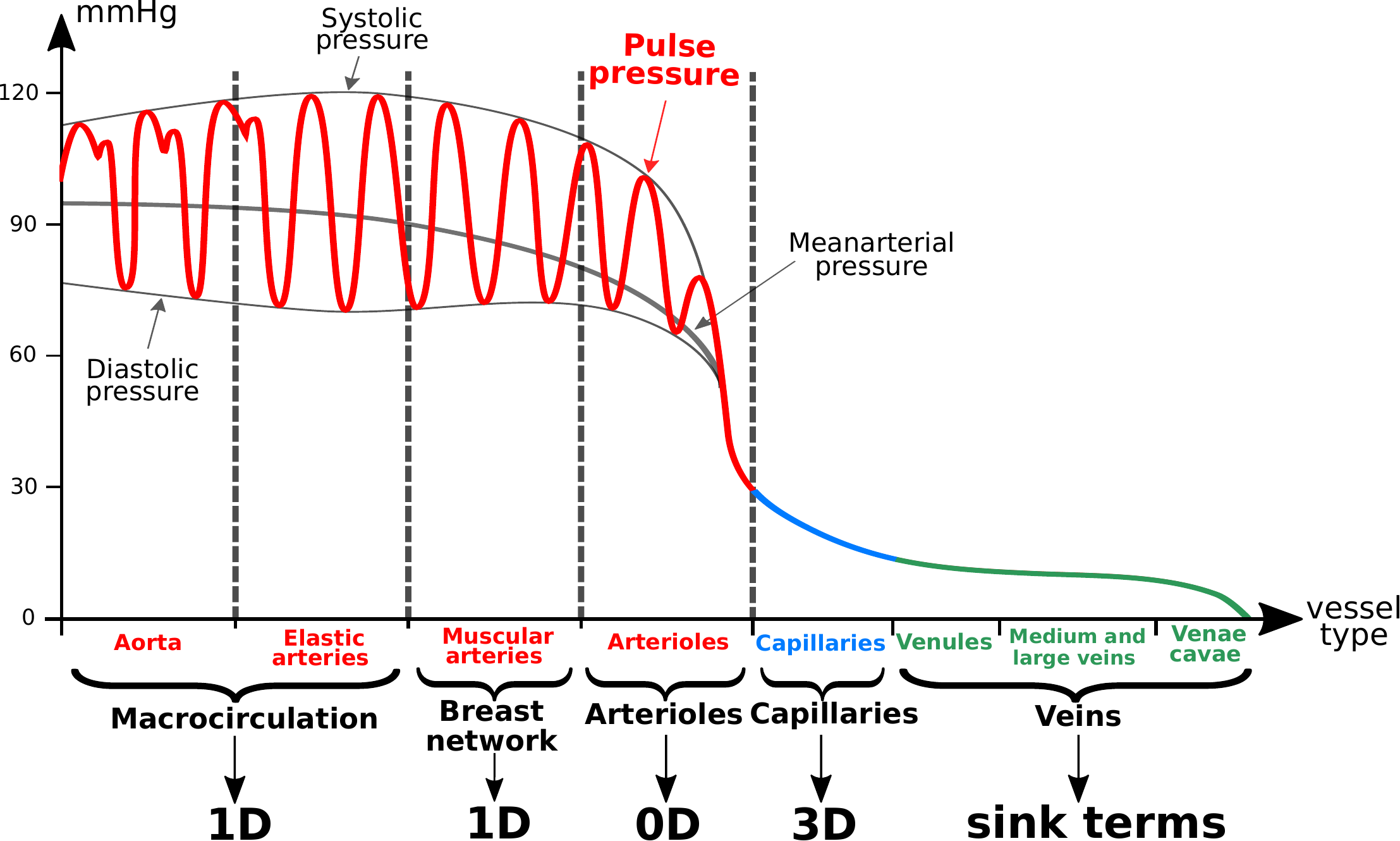}
	\caption{\label{fig:PressureInDiffVessels} Typical pressure curves within various vessel types in the systemic circulation. In our setting, the aorta and the elastic arteries are partially contained in the macrocirculation, while the breast network is composed of muscular arteries. Flow and transport through both networks are governed by means of 1D models. The remainder of the arterial tree is modeled by lumped parameter models (0D models), homogenized 3D models, and sink terms, resulting in a 1D-0D-3D coupled model. The plot of the pulse pressure is based on Fig.~20.10 in Reference~\citenum{biga2020anatomy} (CC BY 4.0).}
\end{figure}

In order to model flow and transport processes from the heart down to the breast tissue, it is necessary to take into account many different vessel types. 
Considering Figure \ref{fig:PressureInDiffVessels}, the systemic vessel tree consists of the aorta, elastic arteries, muscular arteries, arterioles, capillaries, and veins. 
The exchange of fluid and substances between the vasculature and tissue occurs at the capillary level. 

Before the model equations for the different parts of the vascular tree and the tissue are presented, we discuss the most important modeling assumptions and simplifications, which are essential for the design of the mathematical model. In total, $8$ assumptions for our modeling concept are introduced:
\ \\ 
\begin{enumerate}[label=(A\arabic*), ref=A\arabic*, wide, labelindent=0pt] 
	\item \textbf{Blood flow in the macrocirculation is pulsatile and turbulent. Elastic deformation behavior of the vessel walls is assumed.} 
	According to Table~1.7 in Formaggia et al.,\cite{Quarteroni} blood flow in most vessels of the macrocirculation exhibits Reynolds numbers that are larger than $1000$. 
	Therefore, we assume that turbulent flow is present in this sub-network. 
	Furthermore, flow in these vessels is pulsating (see Figure \ref{fig:PressureInDiffVessels}). 
	Since the vessel walls of the larger arteries are highly deformable, a fluid structure interaction (FSI) model has to be considered which couples the flow in the lumen of the vessel with the deformation of the vessel wall. \label{Ass:BloodTurbulent}

    \item \textbf{Blood flow in the breast network and extension is laminar and the deformations of the vessel walls are small.} 
	Considering the range of diameters occurring in the breast network, it can be concluded that the breast network consists mainly of small arteries. 
	The Reynolds numbers occurring in these vessels are significantly lower (see Table~1.7\cite{Quarteroni}) compared to the ones in large arteries. 
	Thus, we can expect laminar flow in these vessels. 
	Moreover, the deformation of the vessel walls is much smaller than that of the larger vessels. \label{Ass:BloodLaminar}

	\item \textbf{Tissue is modeled as a porous medium.}
	Considering the composition of a tissue block, it can be observed that it is mainly composed of cells, fibers, and interstitial space filled with a fluid similar to blood plasma. 
	The interstitial space exhibits several pores that are connected by pore throats. 
	Therefore, it is reasonable to consider tissue as a porous medium which can be described by REV-based flow models\cite{khaled2003role,possenti2019computational,stoverud2012modeling} constructed from Darcy's law.
		
	\item \textbf{Inertial effects concerning flows in the capillary bed and tissue are not considered.} 
	According to Table~1.7,\cite{Quarteroni} blood velocity is about $0.1\;\left[ \unitfrac{mm}{s} \right]$ in the arterioles and venules and about $0.01\;\left[ \unitfrac{mm}{s} \right]$ in the capillary bed in a human system. 
	Therefore, it is reasonable to assume that the Reynolds numbers are significantly lower than $1.0$. 	\label{Ass:TissueNoInertial}

	\item \textbf{Drainage systems are modeled as sink terms.}
	In \eqref{Ass:BloodTurbulent} and \eqref{Ass:BloodLaminar}, we have considered the supply systems for the breast tissue. 
	To remove the fluid mass flowing into the breast tissue, drainage systems are required. 
	Within the systemic circulation, the veins and lymphatic system act as drainage systems.
	As in the case of capillaries and arterioles, there are no data available describing both the venous and lymphatic system. 
	Due to that, we model these drainage systems by means of sink terms within Darcy's equation for both the capillary bed and tissue. 
	Therefore, the sink term representing the veins is assigned to the capillary system, while the source term for the lymphatic system is added to the tissue model. \label{Ass:DrainageSink}
	
	\item \textbf{Gravity effects are neglected.} \label{Ass:GravityNo}
		
	\item \textbf{Blood is an incompressible fluid.} 
		In general, the density of blood is non-constant.\cite{Quarteroni}
	However, in many publications\cite{cattaneo2014fem,erbertseder2012coupled,peyrounette2018multiscale,possenti2019computational,secomb2013angiogenesis} that are concerned with the modeling of blood flow in microvascular networks, compressibility is neglected with only minor effects on the flow.  \label{Ass:BloodIncompressible}

	\item \textbf{The non-Newtonian flow behavior is accounted for by an algebraic relationship.}
	Red blood cells govern the viscosity of blood $\mu_{\mathrm{bl}}\;\left[ \unit{Pa \cdot s} \right]$ significantly, since the red blood cells have to deform when they move through capillaries. 
	We use the relation\cite{pries1996biophysical} which depends on the vessel radius and holds for human blood.\label{Ass:NewtonianNon}
\end{enumerate}

Based on the assumptions \eqref{Ass:BloodTurbulent}--\eqref{Ass:NewtonianNon}, we design in the following sections a class of models for flow and transport in breast tissue and its vasculature.

%% file: multiscale_3_model_components.tex
\section{Multiscale model components}
\label{sec:modelComponents}

In this section,  the model equations for blood flow and transport are introduced, which are based on the assumptions \eqref{Ass:BloodTurbulent}--\eqref{Ass:NewtonianNon} and assigned to different parts of the vasculature.

\subsection{The 1D model}
We use reduced 1D models for the parts of the vasculature for which we have topological information describing the connectivity between vessels with known diameters and lengths. Recalling the description of the data sets in Section \ref{sec:modelSetting}, this is the case for the macrocirculation $I_{macro}$, the thoracic arteries $I_{tho}$ and the breast network $I_{breast}$. The single vessels $\Lambda_i \equiv [0, l_i]$ of length $l_i$ are considered as one-dimensional manifolds, which are glued together at their boundaries, forming a graph structure. For a time-dependent scalar field $F$ on the vasculature we denote its restriction to $\Lambda_i$ by $F_i$ and its evaluation at time $t\geq 0$ at $z \in \Lambda_i$ with $F_i(z,t)$.

\subsubsection{The nonlinear model}\label{sec:nonlinearmodel}
For the larger arteries contained in $I_{macro}$ and $I_{tho,1}$, we have to consider assumptions \eqref{Ass:BloodTurbulent}, \eqref{Ass:BloodIncompressible} and \eqref{Ass:NewtonianNon}.
In order to model blood flow within these vessels, we use a nonlinear hyperbolic PDE model.\cite{formaggia2001coupling,koppl2013reduced,masri2021discontinuous}
It describes time-dependent scalar fields for the flow $Q_i$~$[\si{\cubic\cm\per\s}]$ and vessel area $A_i$~$[\si{\square\cm}]$ on each 1D-vessel $\Lambda_i, i \in I_{non}$, where $I_{non}$ contains the vessel indices on which we impose this model. The resulting velocity fields $Q_i/A_i$ are then coupled to a 1D transport equation for the line concentration field $\Gamma_i$~$[\si{\milli\mole\per\cm}]$.

For vessel $i \in I_{non}$, we denote the backward and forward propagating characteristics of the flow problem by $W_{1,i}$ and $W_{2,i}$, while $W_{3,i}$ is the single, scalar characteristic for the transport problem. These characteristics enable us later to infer coupling conditions. The couplings of nonlinear model at branching points follows the procedure described in Masri et al.\cite{masri2021reduced}

\subsubsection{The linearized model}\label{sec:linearizedmodel}
For smaller vessels further down the vascular tree, i.e, the vessels contained in $I_{tho,2}$, $I_{tho,3}$ and $I_{breast}$, the deformations are less significant. Thus a simplified linearized model can be applied, based on assumptions \eqref{Ass:BloodLaminar}, \eqref{Ass:BloodIncompressible} and \eqref{Ass:NewtonianNon}. We collect the vessel indices for this model in $I_{lin}$.
It describes the pressure $p_i$~$[\unit{Ba}]$ and flow $q_i$~$[\si{\cubic\cm\per\s}]$ fields for each vessel $i \in I_{lin}$. The model itself is given by a 1D hyperbolic PDE system.
A derivation is given in Section 6.2.1 in Reference~\citenum{d2007multiscale}.
The resulting velocities $q_i/A_{i}$ are coupled to the transport of the concentration $\Gamma_i$ as in the nonlinear case. We denote the characteristics for the flow by $w_{1,i}$, $w_{2,i}$ and for the decoupled transport by $w_{3,i}$. Thereby $w_{1,i}$ is the backward traveling wave, while $w_{2,i}$ is the forward traveling wave. The coupling of linear models at branching points using characteristics is discussed in Chapter~7 in Reference~\citenum{d2007multiscale}.

\subsection{The 3D model}
\label{sec:3dmodel}

In this subsection, we consider the lowest level within the vascular tree, i.e., the capillary bed and the tissue matrix. Both systems have to be combined since the capillaries exhibit gaps in their thin vessels such that an exchange between the vascular system and the cells can take place.
In order to avoid a discrete resolution of the capillary network, homogenized Darcy-type models have been investigated in literature.\cite{ehlers2015multi,hodneland2019new,jozsa2021porous,kremheller2021validation,rohan2018modeling,vidotto2019hybrid,levick2010Microvascular}
Following this approach and modeling the lymphatic and venous systems by means of source terms (see Assumption \eqref{Ass:DrainageSink}), one obtains for the capillary bed (3D porous medium, flow problem)
	\begin{equation}
	\label{eq:capillary_system}
	\left\{
	\begin{aligned}
	-\nabla \cdot \left( \rho_{\mathrm{bl}} \frac{\mathbf{K}_{\mathrm{cap}}}{\mu_{\mathrm{c}}} \nabla p_{cap}\right)&= q_{\mathrm{cv}}
	+ q_{\mathrm{ct}} + q_{\mathrm{ca}},&\qquad& \text{in } \Omega,\\
	\rho_{\mathrm{c}} \frac{\mathbf{K}_{\mathrm{cap}}}{\mu_{\mathrm{c}}} \nabla p_{cap} \cdot \mathbf{n} &= 0, &\qquad &\text{on } \partial \Omega,
	\end{aligned}\right.
	\end{equation}
	and for the tissue (3D porous medium, flow problem)
	\begin{equation}
	\label{eq:tissue_system}
	\left\{
	\begin{aligned}
	-\nabla \cdot \left( \rho_{\mathrm{int}} \frac{\mathbf{K}_{\mathrm{t}}}{\mu_{\mathrm{int}}} \nabla p_{\mathrm{t}} \right) &= -q_{\mathrm{ct}} + q_{tl}, &\qquad &\text{in } \Omega,\\
	\rho_{\mathrm{int}} \frac{\mathbf{K}_{\mathrm{t}}}{\mu_{\mathrm{int}}} \nabla p_{\mathrm{t}} \cdot \mathbf{n} &= 0, &\qquad &\text{on } \partial \Omega.
	\end{aligned}\right.
	\end{equation}

\begin{figure}\centering
	\includegraphics[width=0.7\textwidth]{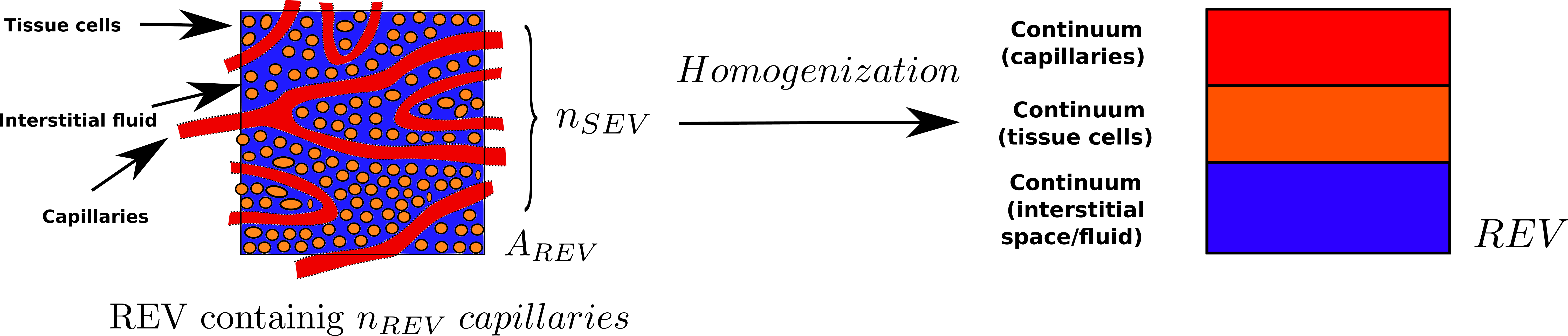}
	\caption{\label{fig:homogenization} Illustration of the capillaries, tissue cells and interstitial fluid, which undergo a homogenization procedure.}
\end{figure}

Here, $\Omega \subset \mathbb{R}^3$ denotes the tissue volume, and $p_{cap}\;\left[\unit{Ba}\right]$ and $p_t\;\left[\unit{Ba}\right]$ represent the pressures in capillary bed and tissue, respectively. 
Moreover, $\rho_{\mathrm{bl}}\;\left[ \unitfrac{g}{cm^3} \right]$ is the density of blood. 
Following J\'{o}zsa et al.,\cite{jozsa2021porous} the diameter of a capillary is between $5.0\;\unit{\mu m}$ and $10.0\;\unit{\mu m}$. 
Thus, we can use $\mu_{\mathrm{bl}}$ from \eqref{Ass:NewtonianNon} with an averaged capillary radius $\overline{r}_c = 3.75\;\unit{\mu m}$ to define the viscosity $\mu_c = \mu_{\mathrm{bl}}\left( \overline{r}_c \right)$.
The permeability tensors $\mathbf{K}_{\mathrm{cap}}\;\left[\unit{cm}^2\right]$ and $\mathbf{K}_{\mathrm{t}}\;\left[\unit{cm}^2\right]$ have the following shape: $\mathbf{K}_{\mathrm{cap}} = k_{\mathrm{cap}} \cdot \mathbf{I}_3$ and $\mathbf{K}_{\mathrm{t}} = k_t \cdot \mathbf{I}_3$, where $\mathbf{I}_3 \in \mathbb{R}^{3 \times 3}$ is the identity matrix. 
In contemporary medical literature,\cite{cattaneo2014fem} a constant value for $k_t$ is chosen. 
Following Reference~\citenum{jozsa2021porous}, the permeability of the capillary bed can be estimated as follows:
\[
k_{\mathrm{cap}} = \frac{n_{SEV} \cdot \overline{r}_c^4 \cdot \pi}{8 \cdot A_{REV}}.
\]
Here, $n_{SEV}$ is the number of capillaries passing a facet of a cubed REV, see Figure \ref{fig:homogenization}, having the area $A_{REV}$. In References~\citenum{kremheller2021validation,vidotto2019hybrid}, the edge length of a cubed REV is about $1.0\;\unit{mm}$. Thus, $A_{REV} \approx 1.0\;\unit{mm}^2$. 

On the right hand side of \eqref{eq:capillary_system}, three different source terms appear. The first one, $q_{cv}$, accounts for the drainage of the capillary bed by means of the venous vessel systems. Following Chapter~6 in Reference~\citenum{d2007multiscale}, it is given by:
\[
q_{cv} = \rho_{bl} \cdot L_{cv} \cdot \left( p_{v} - p_{cap} \right),
\]
where $p_v$ is an averaged venous pressure. $L_{cv}\;\left[ \unitfrac{1}{\unit{Ba} \cdot s} \right]$ represents a conductivity parameter for the blood transfer from the capillary system into the venous system. In a next step, we examine the source term $q_{ct}$ which includes the flow of blood plasma from the capillaries into the interstitial space of the tissue matrix. To determine this source term, a simplified version of Starling's filtration law is considered. Neglecting the oncotic pressures,  it reads as follows:\cite{levick2010Microvascular}
\[
q_{ct} = \rho_{\mathrm{int}} \cdot L_{ct} \cdot S_{ct} \cdot \left[ \left(p_{t} - p_{cap} \right) \right].
\]
The hydraulic conductivity of the vessel wall is given by the parameter $L_{ct}\;\left[ \unitfrac{cm \cdot s}{Ba} \right]$. Another crucial parameter governing the fluid exchange between vascular system and interstitial space is the surface area per REV: $S_{ct}\;\left[ \unitfrac{cm^2}{cm^3} \right]$. This parameter can be estimated similar to $k_{cap}$:
\[
S_{ct} = n_{REV} \cdot 2 \cdot \overline{r}_c \pi \cdot \overline{l}_c/V_{REV} = n_{REV} \cdot S_{cap}/V_{REV}.
\]
Here, $n_{REV}$ denotes the average number of capillaries per REV and $\overline{l}_c$ is the average length of a capillary, i.e., the average surface area of a capillary is given by: $$S_{cap} = 2 \pi \cdot \overline{r}_c \cdot \overline{l}_c.$$ Moreover, $V_{REV}$ denotes the volume of an REV. 
The last term on the right hand side of \eqref{eq:capillary_system} incorporates the blood transfer from the arterioles into the capillaries and is described in Section~\ref{sec:0d-3d-coupling}, since it couples the two models used for these vessel types. 

Finally, it remains to specify the source term for the drainage of the interstitial space via the lymphatic system. Here, we assume it is given by:
\[
q_{tl} = \rho_{\mathrm{int}} \cdot L_{tl} \cdot \left( p_l - p_t \right),
\]
where $p_l= 1333.22\; [\unit{Ba}]$ and $L_{tl}\;\left[ \unitfrac{1}{\unit{Ba} \cdot s}\right]$ is again a conductivity parameter for the lymphatic drainage process.
In order to model the transport processes within the capillary bed and tissue, two volumetric concentration values $c_{cap}\;\left[ \unitfrac{mmol}{cm^3} \right]$ and $c_{t}\;\left[ \unitfrac{mmol}{cm^3} \right]$ are considered. 
Following \eqref{Ass:TissueNoInertial}, the evolution of both concentration variables are governed by stationary convection-diffusion equations. In case of the capillary bed, we obtain:
	\begin{equation}
	\label{eq:capillary_system_transport}
	\left\{
	\begin{aligned}
	\nabla \cdot \left( \mathbf{v}_{cap} \cdot c_{cap} - D_{cap} \nabla c_{cap}\right) &= t_{\mathrm{cv}}
	+ t_{\mathrm{ct}} + t_{\mathrm{ca}},&\qquad& \text{in } \Omega,\\
	\left( \mathbf{v}_{cap} \cdot c_{cap} - D_{cap} \nabla c_{cap}\right) \cdot \mathbf{n} &= 0, &\qquad &\text{on } \partial \Omega,
	\end{aligned}\right.
	\end{equation}
	For tissue the equations read as follows:
	\begin{equation}
	\label{eq:tissue_system_transport}
	\left\{
	\begin{aligned}
	\nabla \cdot \left( \mathbf{v}_{t} \cdot c_{t} - D_{t} \nabla c_{t} \right) &= -t_{\mathrm{ct}} + t_{tl} + r_{t}, &\qquad &\text{in } \Omega,\\
	\left( \mathbf{v}_{t} \cdot c_{t} - D_{t} \nabla c_{t} \right) \cdot \mathbf{n} &= 0, &\qquad &\text{on } \partial \Omega.
	\end{aligned}\right.
	\end{equation}
The velocity fields $\mathbf{v}_{cap}\;\left[ \unitfrac{cm}{s} \right]$ and  $\mathbf{v}_{t}\;\left[ \unitfrac{cm}{s} \right]$ are computed by means of Darcy's law:
\[
\mathbf{v}_{cap} = -\frac{\mathbf{K}_{\mathrm{cap}}}{\mu_{\mathrm{c}}} \nabla p_{cap}, \qquad
\mathbf{v}_{t} = -\frac{\mathbf{K}_{\mathrm{t}}}{\mu_{\mathrm{int}}} \nabla p_{t},
\]
based on the fluid pressure $p_{cap}$ and $p_{t}$ from \eqref{eq:capillary_system} and \eqref{eq:tissue_system}. On the other hand, diffusivity of the transported substance in the capillary and interstitial space is governed by $D_{cap}\;\left[ \unitfrac{cm^2}{s}\right]$ and $D_{t}\;\left[ \unitfrac{cm^2}{s}\right]$, respectively. As in the case of the flow equations, there are in total four different source and sink terms incorporating the transfer of mass between the different compartments of our model. The first sink term $t_{cv}\;\left[ \unitfrac{mmol}{cm^3 \cdot s} \right]$ models the mass flux from the capillary bed into the venous system. Using the flux $q_{cv}$ from \eqref{eq:capillary_system}, the mass transferred from the capillary bed into the venous system, reads as follows:
\[
t_{cv} = q_{cv} \cdot c_{cap}.
\]
As in case of the flow problem, the $t_{ca}$ term describing the mass flux between the arterioles and the capillary system is discussed in Section~\ref{sec:0d-3d-coupling}.
Furthermore, we have for the lymphatic drainage system:
\[
t_{tl} = q_{tl} \cdot c_{t},
\]
and the exchange between capillaries and interstitial space:
\[
t_{ct} =
\begin{cases}
q_{ct} \cdot c_{t},& \text{ if } q_{ct} \geq 0, \\
q_{ct} \cdot c_{cap},&   \text{ if } q_{ct} < 0.
\end{cases}
\]
The only source term that is not related to source and sink terms of the flow problems \eqref{eq:capillary_system} and \eqref{eq:tissue_system} is the reaction term $r_t$. For simplicity, we use in this work a simple decay term:
\[
r_t = - \lambda_t c_t,
\]
where $\lambda_t\;\left[ \unitfrac{1}{s} \right]$ is the corresponding decay rate. 
Depending on the transported substance more sophisticated reaction terms could be taken into account. 
In case of oxygen, e.g., the Michaelis--Menten law\cite{d2007multiscale} could be considered to model metabolism.

\subsection{0D models}
\label{sec:0D-coupling-models}

After modeling flow and transport within the larger arteries as well as the microcirculation contained in the breast, we consider in this subsection the remaining parts of the systemic blood vessel system, i.e., the vessels located beyond the outlets of larger arteries  and the arterioles attached to the breast network.

\subsubsection{The Windkessel model}
\label{sec:windkessel-model}

Considering the network composed of larger arteries (see Figure \ref{fig:vascularTrees}), it becomes obvious that we neglect a huge number of vessels within the lower body, arms, and head. In order to account for the omitted vessels, we use a Windkessel model described by a surrogate RCR circuit\cite{alastruey2008lumped,Quarteroni}.
The total resistance and conductance originate from a calibration procedure described in Alastruey et al.,\cite{alastruey2007modelling} while we take as an input resistance the one of the adjacent 1D vessel to avoid reflections. In addition, the pressure of the veins is required to close the Windkessel model. Recalling \eqref{Ass:DrainageSink}, the pressure associated with the veins is approximated by an average pressure $p_{ven} = 5.0\;\unit{mmHg}$. Therefore $p_{ven}$ is used to incorporate the flow into the veins. We denote the indices of the involved vessels by $I_{wk} \subset I_{non}$.

\subsubsection{The tree model}
\label{sec:tree-model}

Consistent with the flow regimes indicated in Figure \ref{fig:PressureInDiffVessels}, one can see that the pressure within the arterioles decreases significantly and pulsatile flow is transformed into a uniform flow. Since no data is available describing the arterioles between the breast network and the capillary bed, we introduce a surrogate model based on a number of reasonable physical assumptions.

Attaching a single Windkessel model to the outlets of the breast network is not sufficient to model the damping effect of the arterioles. 
 
Another way to model the arterioles can be found in References~\citenum{rohan2018modeling,padmos2021coupling}. In these publications Darcy type equations are used to emulate the influence of the arterioles. However, in this paper, we consider another approach to avoid ambiguity as to how the permeability tensor of the Darcy-type equation has to be chosen such that the pulsatile pressure in the breast network is damped. 

To model the damping effect, we enhance a modeling concept described in \cite{olufsen1999structured}. The key idea of this modeling concept is to replace the omitted vessels by a structured tree model. In the remainder of this subsection, we outline the details of this tree model as well as some enhancements to be able to simulate the transport of a solute.

\begin{figure}\centering
	\includegraphics[width=0.5\textwidth]{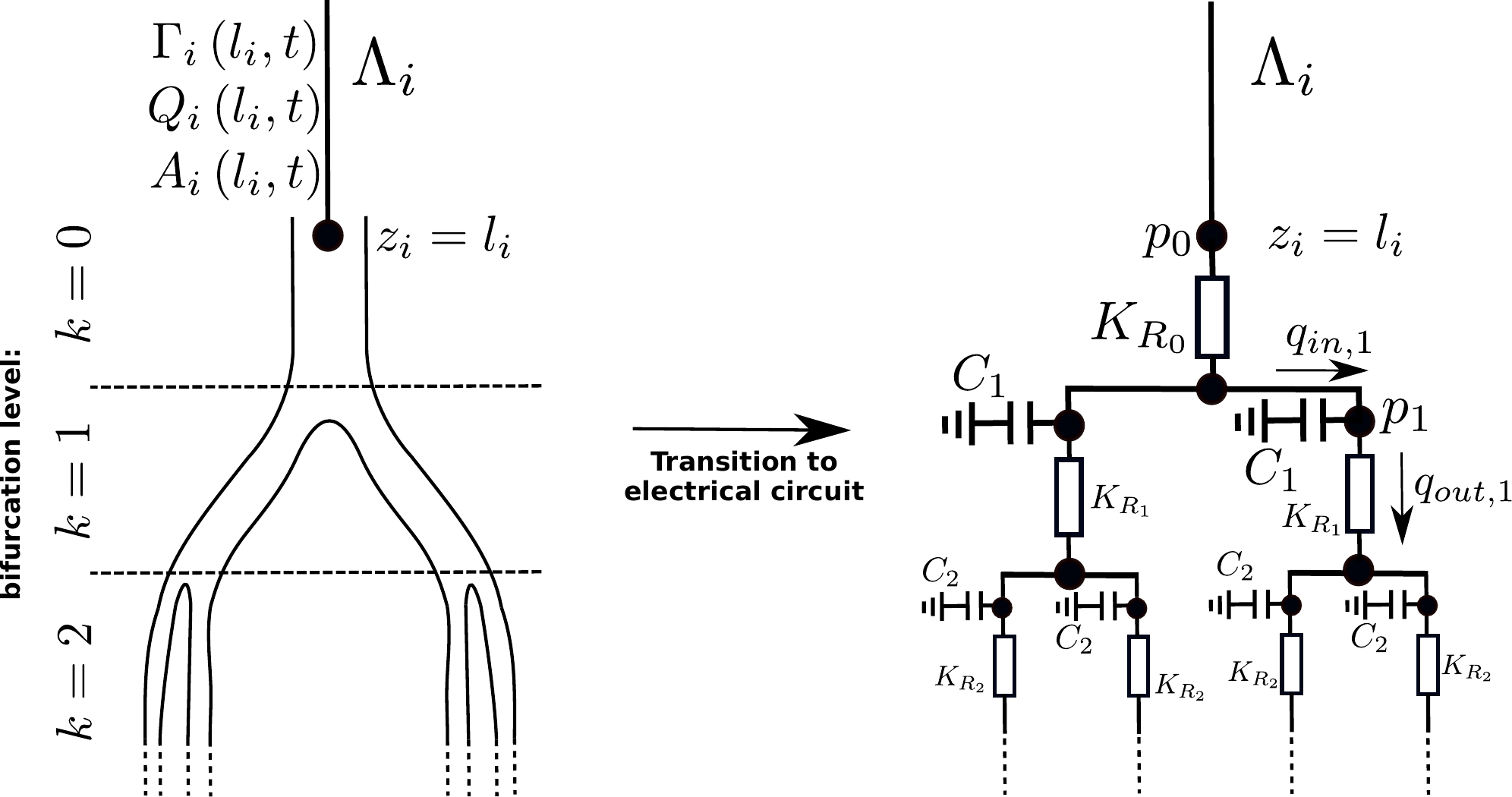}
	\caption{\label{fig:0D_tree} Construction of a surrogate model for the arterioles adjacent to a 1D terminal vessel $\Lambda_i$ using an electrical circuit.}
\end{figure}

All the indices of the vessels to which the ``tree model'' is attached are contained in $I_{tree} \subset I_{lin}$ and we assume that they belong to the terminal vessels of the linearized model.
To design a surrogate model for the missing arterioles, we assume for simplicity that the vessel tree attached to a terminal Vessel $\Lambda_i,\;i \in I_{lin}$ has a symmetric structure. This means that, out of each mother vessel, there are two branches of equal length and radii (see Figure~\ref{fig:0D_tree}). 

The issue arises at to how the lengths and radii of the different vessels can be chosen and how many branching levels should be considered. For this purpose, we use an approach described in References~\citenum{fritz2021modeling,koppl20203d,murray1926physiological,murray1926physiological2,schneider2012tissue,schneider2014tgif}. Let us denote by $r_k$ the radius of the vessels within branching level $k$ and $l_k$ the corresponding length. For $k=0$, we set the radius of the terminal vessel $\Lambda_i$ to $R_i$. According to literature, \cite{murray1926physiological,murray1926physiological2} the radius of a mother vessel $r_m$ and the radii $r_{b1}$ and $r_{b2}$ of the branches are related to each other by the following formula:
\[
    r_m^\gamma = r_{b1}^\gamma + r_{b2}^\gamma,
\]
where $\gamma \in \left[2.5,3.5\right]$ is a parameter that can be tuned. Since we assumed that the vessel tree is symmetric, it follows:
\[
    r_{b1} = r_{b2} \text{ and } r_{b1} = r_m \cdot 2^{-\frac{1}{\gamma}}.
\]
This implies that the radii of branching level $k-1$ and $k$ have the following relationship:
\begin{equation}
    \label{eq:radii_branching}
    r_{k} = r_{k-1} \cdot 2^{-\frac{1}{\gamma}}.
\end{equation}
This recursion also helps us to estimate the number of branching level $N_{B,i}$. Using \eqref{eq:radii_branching} and $r_0 = R_i$, we obtain:
\[
    r_k = R_i \cdot 2^{-\frac{k}{\gamma}}.
\]
In order to estimate $N_{B,i}$, we set $r_k=r_{cap}$, where $r_{cap}$ is the average radius of a capillary. Therefore, one obtains:
\[
    N_{B,i} = -\gamma \cdot \log_2\left( \frac{r_{cap}}{R_i} \right).
\]
Determining the lengths $l_k$, we consider the observation from Schneider et al.\cite{schneider2012tissue,schneider2014tgif} that there is a fixed ratio $ra$ between the length and radius of a vessel. This means that the length $l_k$ is given by: $l_k = ra \cdot r_k$. Here, $ra$ and $r_{cap}$ are given by $28.0$ and $5.0\;\mu m$, see Table~1.7.\cite{Quarteroni}
Since the surrogate vessel tree is fully symmetric, it is sufficient to calculate the pressures only along a single path and not for the entire tree, yielding a linear complexity in the tree-depth. 
For a simplified flow model, the tree is converted into an electrical circuit consisting of resistances $K_{R_k}$ and capacities $C_k$ on each branching level $k$ in the tree hierarchy, where pressure and flow-rates correspond to voltage and current, see Chapter~6.1 in D'Angelo.\cite{d2007multiscale} 
Applying the current-voltage relations for a capacitor on the circuit in Figure~\ref{fig:0D_tree} yields
\[
    C_k \frac{\d p_k}{\d t} = q_{in,k}-q_{out,k},
\]
where $q_{in,k}$ and $q_{out,k}$ are the flows entering and leaving the $k^{\textmd{th}}$ vessel and $p_k$ is the pressure at the capacitor.
These are related by Ohm's law to the pressures such that
\[
    q_{in,k} = \frac{1}{2} \cdot \frac{p_{k-1}-p_k}{K_{R_{k-1}}}, \qquad q_{out,k} = \frac{p_{k}-p_{k+1}}{K_{R_{k}}},
\]
where the factor $\nicefrac{1}{2}$ in $q_{in,k}$ is due to the fact that the flux through resistance $K_{R_{k-1}}$ has to split up equally from the parent vessel into the child vessels (see Figure \ref{fig:0D_tree}). 
Combining the above equations, we obtain
\begin{equation}
    \label{eq:pk}
    C_k \frac{\d p_k}{\d t} = \frac{p_{k-1}-p_k}{2K_{R_{k-1}}}-\frac{p_{k}-p_{k+1}}{K_{R_{k}}},\;\quad k=1,\ldots,N_{B,i},
\end{equation}
where the value $p_{N_{B,i}+1}$ for $k=N_{B,i}$ is determined by a coupling condition (see Section~\ref{sec:0d-3d-coupling}). 
At the other boundary $k=1$, it holds: $p_0 = p_i\left(l_i,t\right)$. 
It remains to provide formulas for the computation of $K_{R_{k}}$ and $C_k$. 
Therefore, we use the expressions from Chapter~6.1 in Reference~\citenum{d2007multiscale}, which are given by
\[
    K_{R_{k}} = \frac{8 \cdot \mu\left( 2\cdot r_k \right) \cdot l_k}{\pi r_k^4}, \qquad C_k = \frac{3 \cdot r_k^3 \cdot \pi \cdot l_k}{2 E_k h_k},
\]
where $E_k$ is the elasticity parameter for the vessels within the $k^{\textmd{th}}$ branching level, $h_k$ represents the thickness of the vessel wall, and the viscosity $\mu$ is again given by \eqref{Ass:NewtonianNon}. 
We set $h_k = \SI{3e-3}{\cm}$ which corresponds to the wall thickness of arterioles listed in Table~1.1.\cite{Quarteroni}
The elasticity parameter was estimated uniformly on all levels as $E_k = \SI{5.2}{\mega\pascal}$. This value is four times higher than the elasticity parameters of the breast network to simulate the damping effect by the arterioles. 
All in all, we determine for each time point a vector of pressures $\mathbf{p} \in \mathbb{R}^{N_{B,i}}$ and flow rates between the different branching levels $\mathbf{q} \in \mathbb{R}^{N_{B,i}+1}$ given by \( q_k = (p_{k}-p_{k+1}) / K_{R_{k}}.  \)

Having the flow field at hand, a model for transport of a solute in blood can be derived. 
In a first step, a concentration value $c_k\;\left[ \unitfrac{mmol}{cm^3} \right]$ and a volume $V_k\;\left[ \unit{cm}^3 \right]$ is assigned to each branching level. 
To determine these variables, for each branching level a mass balance equation for the solute and blood volume can be established. In case of the $k$-th level the equations read as follows:
\begin{align}
    \nonumber
    \frac{\d V_k}{\d t} & = \nicefrac{1}{2}\, q_{k-1} - q_k \quad \textmd{ and } \\
    \frac{\d}{\d t}\left( V_k\left( t \right) c_k\left( t \right) \right) & = N_{k,in} - N_{k,out}
\end{align}
where $N_{k,in}$ and $N_{k,out}$ are the number of particles in $\si{\milli\mole}$ entering and leaving the $k^{\textmd{th}}$ level per second. Using the flow rates and concentrations of the neighboring vessels we get
\begin{align}
    \frac{\d}{\d t}\left( V_k\left( t \right) c_k\left( t \right) \right) & =
    \begin{cases}
        \nicefrac{1}{2}\, q_{k-1} c_{k-1} - q_k c_k, &\text{ if } q_{k-1} \geq 0,\;q_k \geq 0,  \\
        \nicefrac{1}{2}\, q_{k-1} c_{k-1} - q_k c_{k+1}, &\text{ if } q_{k-1} \geq 0,\;q_k < 0, \\
        \nicefrac{1}{2}\, q_{k-1} c_k - q_k c_{k+1}, &\text{ if } q_{k-1} < 0,\;q_k < 0,        \\
        \nicefrac{1}{2}\, q_{k-1} c_k - q_k c_k, &\text{ if } q_{k-1} < 0,\;q_k \geq 0.
    \end{cases}
\end{align}
In case of $k=N_{B,i}+1$, the boundary concentration $c_{N_{B,i}+1}$ is determined by the coupling conditions in Section~\ref{sec:0d-3d-coupling}. 
For $k=0$, $c_0$ is given by the solution in the 1D terminal vessel $\Lambda_i$: $c_0 = \Gamma_i\left(l_i,t\right)/A_i\left(l_i,t\right)$.
In total, we have to determine the two vectors $\mathbf{V} \in \mathbb{R}^{N_{B,i}}$ and $\mathbf{c} \in \mathbb{R}^{N_{B,i}}$.

%% file: multiscale_4_coupling_and_boundary_conditions.tex
\section{Coupling and boundary conditions}
\label{sec:coupling-and-boundary-conditions}

In the previous section, model equations for the different parts of the vascular tree are discussed. It remains to specify the missing boundary and coupling conditions if one of these models is defined on a vessel with an inlet or an outlet. 

First, the inflow boundary conditions for the nonlinear model are introduced. For the linearized model, we prescribe a pressure boundary condition at the inlets. The vessel indices at which an inflow boundary condition is applied are contained in the set $I_{in}$. The boundary conditions and the coupling conditions between the different model components remain to be determined. In Section~\ref{sec:0D-coupling-models}, the coupling of the Windkessel- and Tree-model with the 1D models in the terminal vessels is outlined. The missing coupling conditions between the nonlinear and linearized models are given in Section~\ref{sec:direct-nonlinear-to-linearized-model-coupling}.
For this, the index sets $I^{coup}_{lin} \subset I_{lin}$ and $I^{coup}_{non} \subset I_{non}$ collecting all vessel indices at the interface between the nonlinear and linear model are introduced.
Section~\ref{sec:0d-3d-coupling} discusses the 0D-3D coupling conditions, of the 0D Tree-models for the arterioles with the 3D porous media models for the capillaries.

\subsection{Inflow nonlinear model}
\label{sec:inflow-nonlinear-model}
Let $\Lambda_i$ be an inflow vessel in the nonlinear regime, i.e. $i \in I_{in} \cap I_{non}$.
For convenience, we assume that the parametrization of the vessel $\Lambda_i$ is oriented in such a way that $z_i=0$ is adjacent to the inlet of the aorta. In order to simulate the impact of the heart beats, we prescribe at $z_i=0$ the following profile for the flow rate:
\begin{equation}
\label{eq:Qheart}
Q_{heart}\left( t \right) = Q_{\max} \cdot
\begin{cases}
\sin\left( \frac{t \cdot \pi}{0.3 \cdot T} \right),\; 0 \leq t \leq 0.3 \cdot T, \\
\\
0.0, \; 0.3 \cdot T < t \leq T.
\end{cases}
\end{equation}
$Q_{\max}\;\left[\unitfrac{cm^3}{s}\right]$ is the maximal flow rate and $T\;\left[\unit{s}\right]$ is the duration of the heart beat, which are for our simulations chosen as $Q = \SI{485}{\cubic \cm \per \s}$ and $T = \SI{1}{\s}$.
To determine the flow rate for time points $T < t$, we extend $Q_{heart}$ periodically. 
The remaining boundary condition is obtained from the outgoing characteristic variable $W_{1,i}\left(0,t\right)$. 
By this, the boundary conditions for $A_i$ and $Q_i$ at the inlet can be computed. More details on computing the boundary data at an inflow boundary can be found in literature.\cite{formaggia2002one,masri2021discontinuous} For the concentration variable, we use a constant concentration value $c_{in} = \SI{1}{\milli\mole \per \cubic \cm}$, see Chapter~2 in D'Angelo.\cite{d2007multiscale}
Thus, we have
\begin{equation}
\label{eq:cin}
    \Gamma_i\left(0,t\right) = c_{in} \cdot A_i\left(0,t\right),
\end{equation}   
if $Q_i\left(0,t\right) >0$. In the other case, an upwinding with respect to the concentration variable has to be performed.

\subsection{Inflow linearized model}
\label{sec:inflow-linearized-model}
For each vessel $\Lambda_i$ with $i \in I_{in} \cap I_{lin}$, we assume that a time-dependent pressure profile $\tilde p_i (t)\;\left[\si{\mmHg}\right]$ is given.
If we assume again for convenience that $z_i = 0$ is adjacent to the inflow boundary, then the boundary conditions for flow can be determined from the outgoing characteristic $w_{1,i}(0, t)$ and the pressure profile $\tilde p_i(t)$.

\subsection{Direct nonlinear to linearized model coupling}
\label{sec:direct-nonlinear-to-linearized-model-coupling}

Let us consider two vessels $\Lambda_j,\;j \in I^{coup}_{non}$ and $\Lambda_i,\;i \in I^{coup}_{lin}$ where $\Lambda_i$ and $\Lambda_j$ are connected and should be coupled.
Assuming that $z_j = l_j$ and $z_i = 0$, the outgoing non-linear and linear characteristic variables $W_{2,i}\left(l_j,t\right)$ and $w_{1,j}\left(0,t\right)$ are known, e.g., by extrapolation, \cite{formaggia2003one} which yields two coupling conditions.
In addition, we enforce the conservation of mass and the continuity of pressure:
\begin{equation}
    \label{eq:IT_3}
    Q_j\left(l_j,t\right) = q_i\left(0,t\right), \qquad P_j\left(l_j,t\right) = p_i\left(0,t\right).
\end{equation}
These equations form a non-linear system of equations for the four boundary values $Q_j\left(l_j,t\right)$, $A_j\left(l_j,t\right)$, $p_i\left(0,t\right)$ and $q_i\left(0,t\right)$ that has to be solved for time points of interest. By means of the fluid variables, we can determine the concentration variables $\Gamma_j\left(l_j,t\right)$ and $\Gamma_i\left(0,t\right)$. If the fluid is leaving the respective vessel an upwinding procedure is performed; otherwise, the conservation of solute mass transported through the interface is enforced.

\subsection{Coupling between 0D tree and 3D capillaries}
\label{sec:0d-3d-coupling}
The 0D tree model and the 3D capillary model are coupled by suitable source terms $q_{ca}$ and $t_{ca}$. In order to define these source terms, we decompose the breast volume into perfusion zones. A similar idea\cite{di2021computational} has been considered in context of modeling the myocardial perfusion.

\subsubsection{Perfusion zones}
\begin{figure} \centering
	\includegraphics[width=0.5\textwidth]{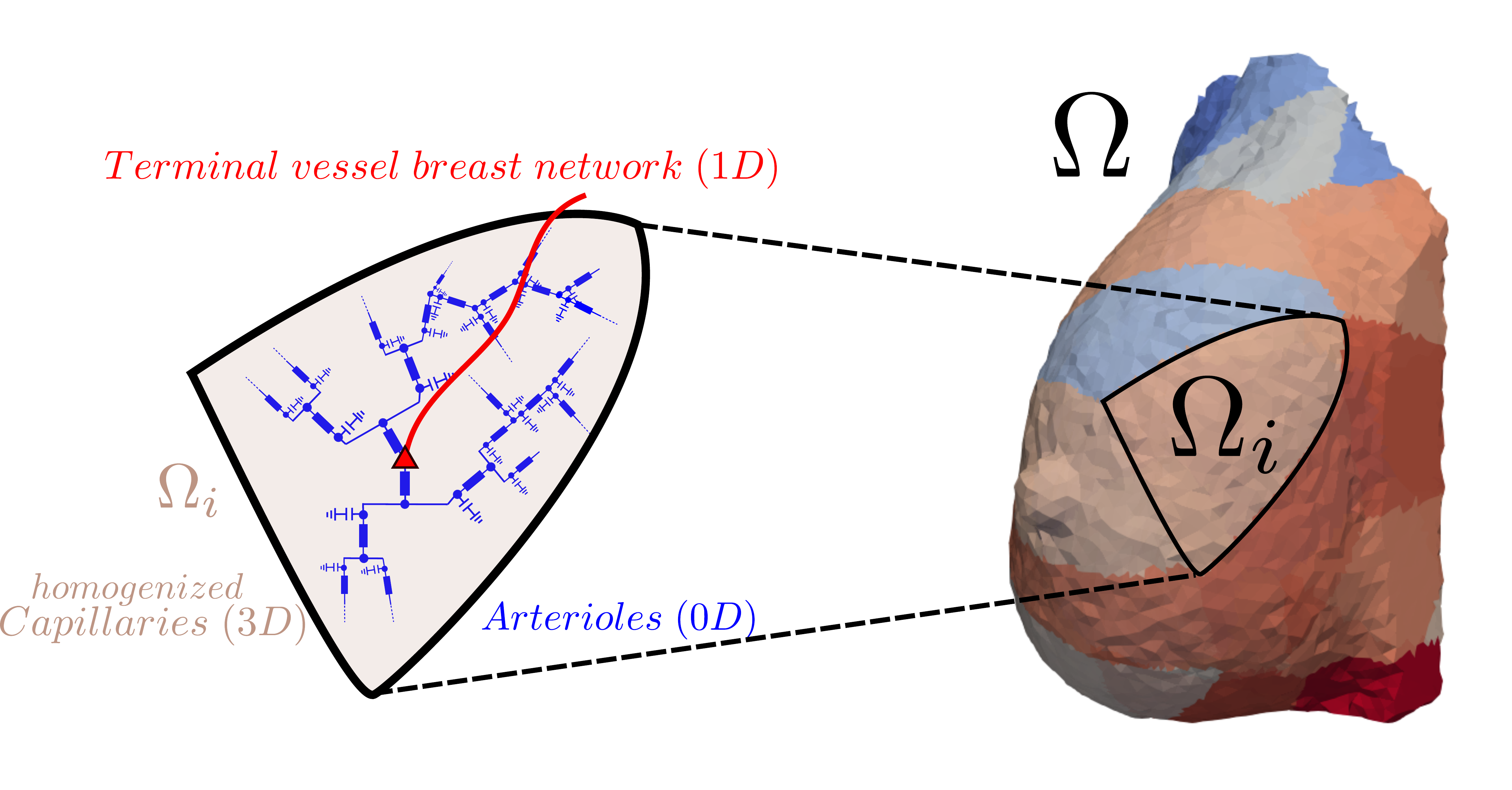}%
	\caption{\label{fig:coupling3d0dtree} 
		Left: Terminal 1D vessel of the breast network supplying the capillary bed contained in a perfusion territory. The arterioles connecting the 1D terminal vessel and the capillaries are modeled by a $0$D model (see Section \ref{sec:0D-coupling-models}). Right: Perfusion zones of the 3D breast geometry with each zone containing an outlet of the 1D geometry.
	}
\end{figure}

To define so-called perfusion zones, the breast volume $\Omega$ is decomposed in $N_{art} = \left|I^{out}_{lin}\right|$ perfusion volumes which correspond to the number of outlets contained in the linearized model. We denote the perfusion volumes by $\Omega_i \subset \Omega$ and assume that they form a disjoint decomposition of our 3D-domain, i.e.,
$
\Omega = \cup_{i \in I^{out}_{lin}} \Omega_i.
$
This decomposition is motivated by the fact that each terminal vessel of the breast network supplies blood to a certain tissue volume. 
Thereby, we assume that the arterioles branching out of a terminal vessel $i \in I^{out}_{lin}$ are distributed in $\Omega_i$ connecting the breast network and the capillary bed (see Figure \ref{fig:coupling3d0dtree}, left). 

Now the challenge arises, how the perfusion areas $\Omega_i$ can be defined. 
In context of our modeling approach, we consider the end point $\mathbf{x}_i \in \Omega$ of a terminal vessel $i$. 
Using this notation, $\Omega_i$ is defined as follows:
\[
\Omega_i = \left\{ \left. \mathbf{x} \in \Omega \;\right| \;\text{dist}\left(\mathbf{x},\mathbf{x}_i \right) < \text{dist}\left(\mathbf{x},\mathbf{x}_j \right),\;\forall j \neq i,\;j \in I_{lin}^{out} \right\}.
\]
Here $\text{dist}\left(\mathbf{x},\mathbf{y}\right)$ is the Euclidean distance between two points $\mathbf{x},\mathbf{y} \in \Omega$. 
This definition is motivated by the assumption that a point in the tissue domain is usually supplied by the terminal vessel with minimal distance to this point. 
The decomposition of $\Omega$ into the different perfusion areas is shown on the right of Figure~\ref{fig:coupling3d0dtree}. 

\subsubsection{0D to 3D coupling}
With respect to each perfusion area the source term $q_{ca}$ is given by:
\[
\left. q_{ca} \right|_{\Omega_i} = \rho_{bl} \cdot L_{ca} \cdot \left( p_{N_{B,i}} -\overline{p}_{cap,i} \right).
\]
$p_{N_{B,i}}$ is the pressure at the tips of the $0D$-tree emulating the influence of the arterioles, while $\overline{p}_{cap,i}$ is the averaged capillary pressure with respect to the perfusion area $\Omega_i$:
\[
\overline{p}_{cap,i} = \frac{1}{\left|\Omega_i \right|} \int_{\Omega_i} p_{cap}\left( \mathbf{x} \right)\;\d\mathbf{x}.
\]
As in the previous source terms, the pressure difference is weighted by a conductivity parameter $L_{ca}\;\left[ \unitfrac{1}{\unit{Ba} \cdot s}\right]$. The mass flux from the arterioles into the capillary system is given by:
\[
\left. t_{ca} \right|_{\Omega_i} = \left. q_{ca} \right|_{\Omega_i} \cdot c_{N_{B,i}},
\]
where $c_{N_{B,i}}$ is the concentration of the last compartment within the 0D tree modeling the arterioles attached to outlet $i$. 

\subsubsection{3D to 0D coupling}
The back coupling from 3D to 0D is done by setting the boundary pressure $p_{N_{B,i}+1}$ and boundary concentration $c_{N_{B,i}+1}$ at the outlets of the $0$D-trees to the averaged values $\bar{p}_{cap,i}$ and $\bar{c}_{cap,i}$ inside perfusion zone $\Omega_i$, i.e.,
\[
    p_{N_{B,i}+1} := \bar{p}_{cap,i} , \quad
    c_{N_{B,i}+1} := \bar{c}_{cap,i} = \frac{1}{\left| \Omega_i \right|} \int_{\Omega_i} c_{cap}\left(\mathbf{x} \right)\;d\mathbf{x} 
    \quad \textmd{ for all } i \in I^{out}_{lin}.
\]

%% file: multiscale_5_coupled_models.tex
\section{Coupled models}
\label{sec:coupledModels}
After introducing model equations for the different parts of the vascular tree in Section~\ref{sec:modelComponents} and establishing suitable coupling and boundary conditions in Section~\ref{sec:coupling-and-boundary-conditions}, we use them to define two different models for flow and transport processes. The first model is referred to as the fully-coupled model (see Section~\ref{sec:fullyCoupledModel}). It covers the whole flow path from the aorta down to the capillaries and tissue of the breast. After that a sub-model of the fully coupled is defined in Section~\ref{sec:oneDirectionalCoupledModel} in which only the vascular tree and tissue in the breast are taken into account. In the remainder of this work, it is referred to as the breast model.

\subsection{Fully coupled model}
\label{sec:fullyCoupledModel}

We assign the nonlinear model to all the larger arteries as well as to the first two vessels Lt I and It I of our extension. Flow and transport in all the other vessels are governed by the linearized model. We thus have
\[ 
I_{non} = I_{macro} \cup I_{tho,1} ,
\qquad
I_{lin} = I_{breast} \cup I_{tho,2} \cup I_{tho,3}
.
\]
Conditions \eqref{eq:Qheart} are imposed at the inlet of the aorta, i.e., we set $I_{in} = I^{in}_{macro}$.
At all outlets of the larger arteries Windkessel models are providing the missing boundary data. Therefore, we have: $I_{wk} = I^{out}_{macro}$. The nonlinear and linearized models are coupled with the coupling conditions introduced in Section~\ref{sec:direct-nonlinear-to-linearized-model-coupling}.
Finally, the linear models in $I^{out}_{breast}$, the $0$D-Tree models and the $3$D models for capillaries and tissue are linked together using the coupling conditions from Section~\ref{sec:direct-nonlinear-to-linearized-model-coupling} and~\ref{sec:0d-3d-coupling}.

\subsection{Breast model}
\label{sec:oneDirectionalCoupledModel}

To design our breast model, we proceed in two steps: First, we need boundary conditions, which are provided by the nonlinear purely generic model in Section~\ref{sec:macrocirculationModel}. In the following, this model is called the macrocirculation model. Then, we simulate in a second step (in Section~\ref{sec:mesoMicroCirculationModel}) flow and transport processes in our patient-specific breast network using the linearized PDE model without the generic components. Due to the fact that this model is focused on flow and transport through middle-sized vessels and microcirculation, we refer to it as a meso-microcirculation model.

The first model is based completely on patient-independent data, thus its results can be precalculated and used for all patients. Obviously, not every patient fits the generic data described before, e.g., the anterior communicating artery (Vessel $31$ in Figure~\ref{fig:vascularTrees}) is not always contained in the Circle of Willis. This could be solved by dividing patients into groups sharing similar networks. For each group a generic network could be derived for which the macrocirculation model is precomputed. A new patient would just have to be assigned to one of these groups, before simulations with respect to the individual breast geometry can be taken into consideration. A further advantage would be that the model for the breast is purely linear. 
Thus, the computational costs are much lower compared to the fully nonlinear model.

\subsubsection{Macrocirculation model}
\label{sec:macrocirculationModel}

To obtain appropriate pressure boundary conditions for the next purely linear model, we only want to simulate the nonlinear model and thus impose 
\[ 
I_{non} = I_{macro} \cup I_{tho,1},
\qquad
I_{lin} = \emptyset 
.
\]
To close the model, we assign Windkessel models to both vessels from the extension network, i.e., $I_{wk} = I^{out}_{macro} \cup I_{tho,1}$.
To calibrate the models, we use the algorithm described in Reference~\citenum{alastruey2007modelling} with one modification: The compliance and resistance for the right auxiliary artery will distributed between itself, the Lt I and It I vessels. The distribution of the capacities is proportional to the fluid masses leaving these three vessels, while the resistances are distributed in a reciprocal way.

We expect that flows predicted by this nonlinear model will reach a steady-state sooner than the fully coupled model. When the pressure and velocity waves become periodic, we can compute the pressures at the outlets $i \in I_{tho,1}$ for one period and extend them periodically. The tip pressure at vessel $i\in I_{tho,1}$ defines a boundary condition $\tilde p_j$ for the adjacent vessel $j\in I_{tho,2}$.

\subsubsection{Meso-microcirculation model}
\label{sec:mesoMicroCirculationModel}

If our data is restricted  to the patient specific parts, we obtain the index sets
\[ 
I_{non} = \emptyset, \qquad
I_{lin} = I_{breast} \cup I_{tho,2} \cup I_{tho,3}
.
\]
As inflow conditions, we prescribe the pressures at the vessels
\[ I_{in} = I_{tho,2} \subset I_{tho} \]
as outlined in Section~\ref{sec:inflow-linearized-model}, with the pressure profiles $\tilde p_i(t), i \in I_{in}$ defined in the previous section.
As in case of the fully coupled model, we combine the 3D and 1D equations using the coupling conditions in Section~\ref{sec:0d-3d-coupling}.

%% file: multiscale_6_numerical_solution_methods.tex
\section{Numerical algorithm}
\label{sec:numsol}

Both 1D flow models and the transport model are discretized by a third order discontinuous Galerkin discretization in space. 
The upwinding is determined from the characteristics which can be calculated analytically in our regime.

\subsection{1D nonlinear flow solver}
In time, we use for the nonlinear flow model an explicit Euler scheme.
The explicit time integrator results in a restriction on the time step sizes due to the Courant--Friedrichs--Lewy (CFL) condition.
In case of our application area this is acceptable, since we need small time steps to resolve the pressure and velocity waves. The algorithm is implemented in a matrix-free way using the GMM++ \cite{renard2020getfem} and Eigen \cite{eigenweb} libraries.
For inter process communication, we use the standard MPI primitives.

\subsection{1D linearized flow solver}
The linearized 1D flow model together with the 0D-Tree model are solved numerically using  an implicit Euler scheme in time.
Thus, we have no restrictions on the time step size in these regimes.
This is convenient in our case, since especially at the beginning of the simulation huge flows from the 3D domain pass the 0D-Tree models and would lead to a severe restriction on the time step size for the 1D model. 
The implicit-linearized scheme can now act as a buffer between the explicit-nonlinear and the 3D domain making the whole scheme more robust.

For inverting the 1D-linear system, we use the PETSc library,\cite{petsc-user-ref}
with GMRES as an outer solver and a per processor ILU factorization as a preconditioner.
Since all the degrees-of-freedom on a blood vessel get distributed as a whole to one processor the ILU is expected to work well in this case.
In addition, since the underlying matrix does not change over time, the factorization has only to be calculated once. 

\subsection{1D-1D coupling}
The fully coupled model from Section~\ref{sec:fullyCoupledModel} needs a direct coupling between the nonlinear and linearized regimes. Thus, both schemes have to work on the same time scale. 
Their coupling in time is depicted in Figure~\ref{fig:implExplCharacteristicCoupling}.
We first advance the nonlinear scheme with the old characteristic boundary values.
The new values from the nonlinear scheme are then fed into the characteristic boundary conditions of the implicit linearized scheme. Using the linearized characteristic as input yields a linear problem, which can be inverted by a standard solver, while using the nonlinear characteristics would result in a nonlinear boundary condition. This requires the usage of a nonlinear solver.

Note that with this coupling the explicit nonlinear solver still restricts the time step size of the linearized solver.
This is not the case with the model introduced in Section~\ref{sec:mesoMicroCirculationModel} whose 1D discretization is purely implicit and thus exhibits no time step restriction.

\subsection{0D-3D coupling}
\begin{figure} \centering
	\includegraphics[width=.38\textwidth]{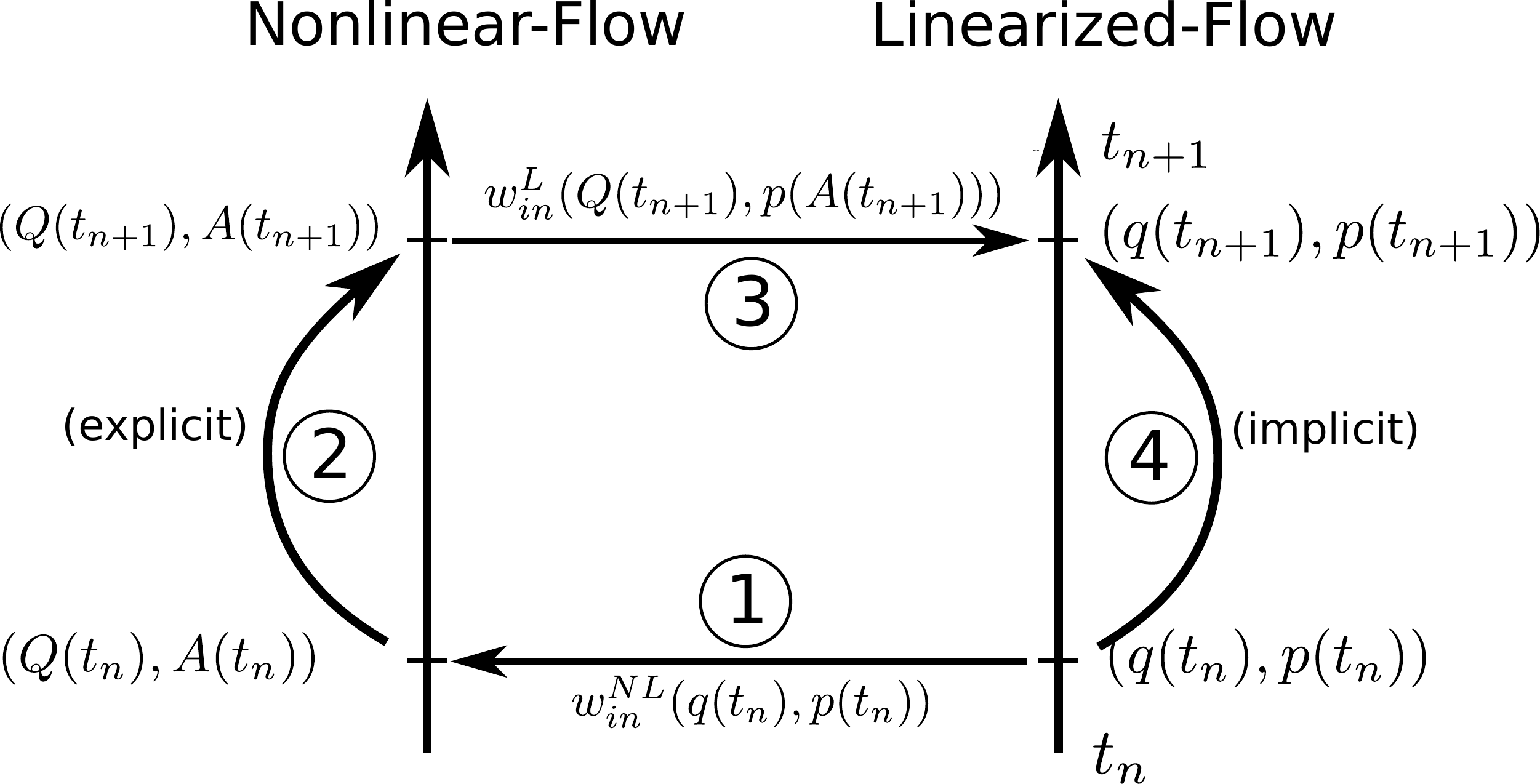}  \qquad
	\includegraphics[width=0.35\textwidth]{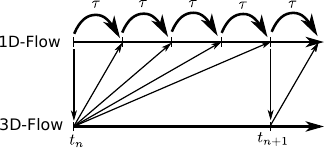}
	\caption{\label{fig:multiRateTimeStepping}\label{fig:implExplCharacteristicCoupling} Left: Coupling of the nonlinear-explicit and linearized-implicit 1D schemes in time.
		Here, $w^L_{in}$ and $w^{NL}_{in}$ represent the incoming characteristics of the linearized and nonlinear models. Right: A simple multi-rate time stepping scheme. After several 1D time-steps, the boundary conditions of the 3D Darcy equation are updated and evaluated. 
		The new average 3D pressures at the vessel outlets are used as new boundary conditions for the linearized 1D flow.}
\end{figure}

For the spatial discretization of the 3D capillary flow problem, we use the space of continuous piecewise-linear polynomials as ansatz functions. 
The matrices are assembled in FEniCS. To retain the sparsity of the system one additional scalar variable per compartment is introduced which holds the pressure to an average value.
The coupling between 0D and 3D is depicted in Figure~\ref{fig:multiRateTimeStepping}.
Due to the small time steps used in terms of the 1D systems, the 3D system is not solved in every time step. Hence, we apply a very basic multi-rate time stepping scheme, which only update the 3D pressures after the time period $\tau_{3D}$ and then update the 1D boundary condition.
This crude coupling works, due to the small changes of the averaged 3D-pressure in time. 
We solve the 1D systems until the solutions become periodic at a time point $t_{\textmd{init,1D}}$. After that, we activate the 0D-3D coupling and simulate until the flow simulation shows a periodic behavior.

\subsection{Transport solvers}
For a time integrator for the 3D, 1D and 0D models, we use the implicit Euler method. 
Since the transport processes in the breast capillaries and tissue are slower compared to the transport process in the vascular tree, we use in case of the 3D models a larger time step size, which aligns with the one for the flow problems. 
The 3D and 0D transport models are coupled in the same manner as the flow solver. 
Discontinuous Galerkin methods are used to discretize the 1D model equations in space, while standard linear finite elements are considered for the 3D model equations. 
The discontinuous Galerkin method has to be combined with a slope limiter technique in order to avoid the formation of spurious oscillations in the vicinity of steep gradients.\cite{kuzmin2010vertex,kuzmin2014hierarchical}

%% file: multiscale_7_results.tex
\section{Results and discussion}
\label{sec:results}

The results obtained using our fully coupled model are first presented and then compared to available biological data by moving through the vessel hierarchy, starting at the inlet of the aorta and ending at the homogenized capillaries. Since the fully coupled model is considered to be our reference model, the results are compared to the simpler breast model, in particular the meso-microcirculation model. This allows us to assess the quality of this sub-model. The source code underlying all numerical experiments, including data for testing, is available at: 

\begin{center} \url{https://github.com/CancerModeling/Flows1D0D3D}
	\end{center} \medskip

As model parameters, we use $\rho_c = \SI{0.997}{\gram\per\cubic\cm}$ and $\rho_t = \SI{1.060}{\gram\per\cubic\cm}$ for the densities of blood in the capillaries and tissue.
We estimate $S_{\textmd{ct}}$ by assuming $n_{\textmd{REV}} = 460$ which is a lower estimate for the capillary density in muscles,\cite{mcguire2003estimation}
$\bar r_c = \SI{3.375e-4}{\cm}$ and $\bar l_c = \SI{0.06}{\cm}$.
We set $k_{cap} = \SI{1e-9}{\cm\squared}$ and $k_{t} = \SI{1e-14}{\cm\squared}$ from Table~1 in Vidotto et al.\cite{vidotto2019hybrid}
For the mean vene and lymphatic pressures we assume $p_{\textmd{v}} = \SI{10}{\mmHg}$ and $p_{\textmd{l}} = \SI{1}{\mmHg}$.
Applying Ohm's law to the last compartment of the 0D model and averaging yields  
$L_{ca} = \frac{\rho_{bl}}{|I^{out}_{breast}|} \sum_{i\in I^{out}_{breast}} \frac{1}{|\Omega_i|} \frac{2^{N_{B,i}}}{K_{R_{N_{Bi,i}}}} $
as an estimate for $L_{ca}$.
The remaining permeabilities are set to $L_{cv} = L_{tl} = 10^{-8}\; \unit{Ba}^{-1} s^{-1} $ and $L_{ct} = 10^{-9}\; \textmd{cm}\,\textmd{s}\,\unit{Ba}^{-1}$.
For the transport we use the oxygen specific values $D_t = \SI{1.7e-5}{\cubic \cm \per \s}$ and $\lambda_t = \SI{6e-5}{\per\s} $ from Reference~\citenum{d2007multiscale} and set $D_c = D_t$.

The numerical parameters for the fully coupled model are set to $\tau = 2^{-16} \si{\s}$, $t_{init,1D} = \SI{6}{\s}$ and $\tau_{3D} = 2^{-3} \si{\s}$.
The boundary condition for the transport is activated after $\SI{6}{\s}$.

\subsection{Fully coupled model}
\label{sec:resultsFullyCoupledModel}

\begin{figure} \centering
	\includegraphics[width=0.56\textwidth]{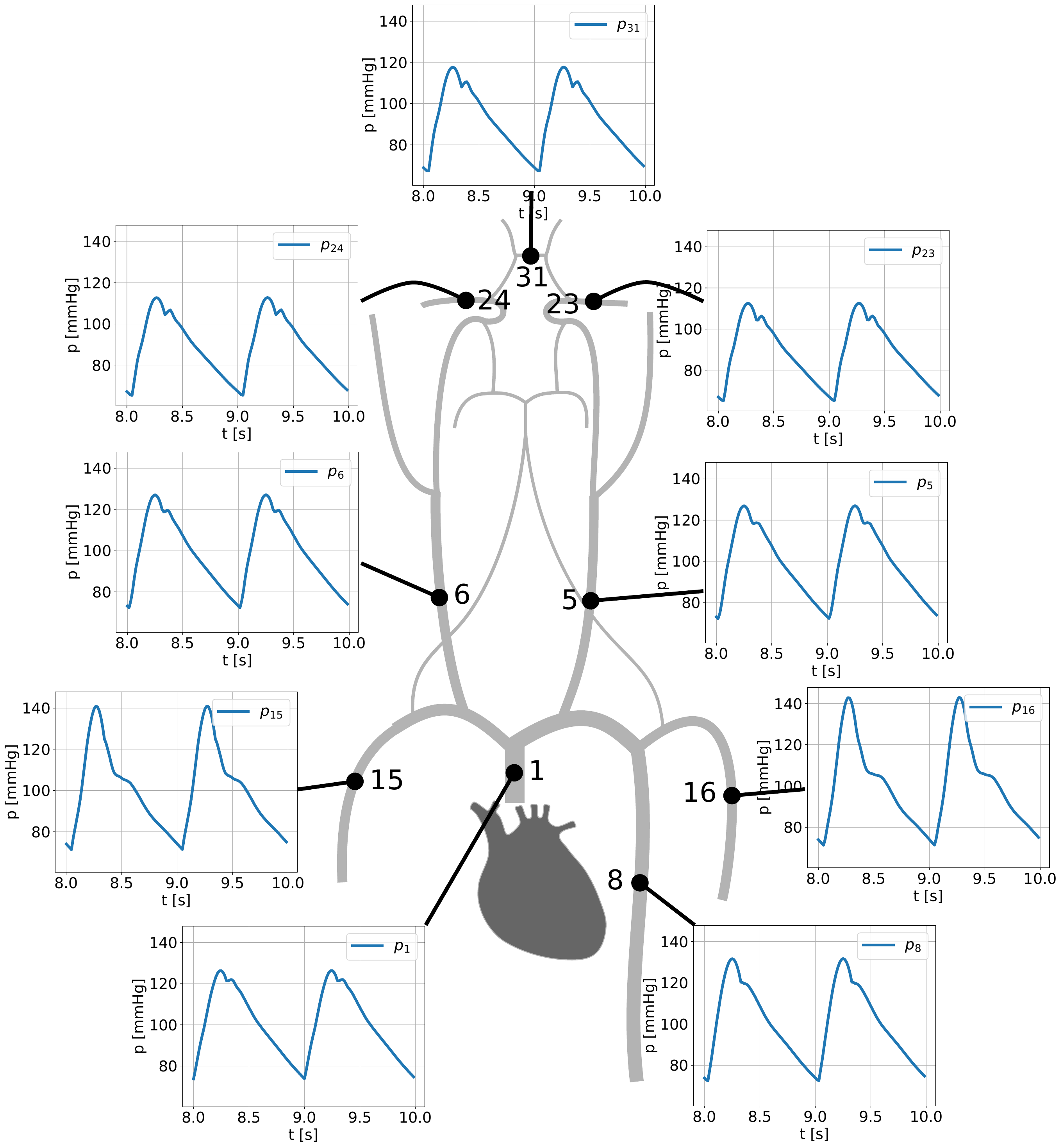}
	\vspace{0.5cm}
	\includegraphics[width=0.56\textwidth]{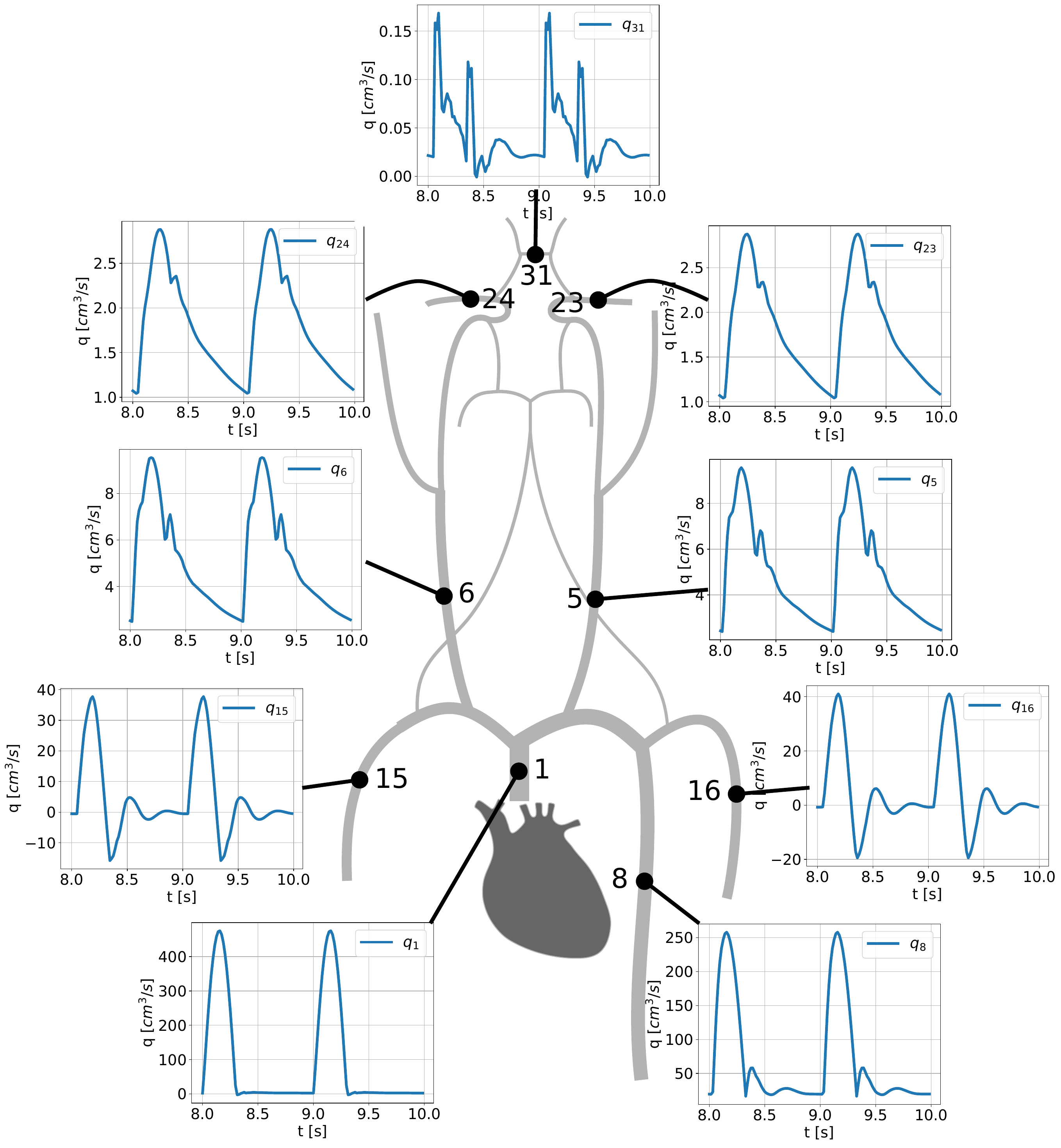}%
	\caption{\label{fig:results1DCoW}
		Pressures [\si{\mmHg}] and flows $\left[ \unitfrac{cm^3}{s} \right]$ at vessel midpoints in the nonlinear regime around the Circle of Willis from \SI{14}{\s} to \SI{16}{\s}.	}
\end{figure}

Results related to the 1D flow models are presented first. In Figure~\ref{fig:results1DCoW}, the pressures and flow rates in the nonlinear regime of our model are depicted, which contains the circle of Willis.

This scenario describes a good benchmark test for the nonlinear part of our model. The pressure at the entrance of the aorta is \SI{130/80}{\mmHg}, while the pressures in the arm arteries are a little higher at \SI{140/80}{\mmHg}, both of which are well within a physiologically meaningful regime. While the order of magnitude of pressure values stays constant in the whole nonlinear regime, the same is not true for the flows, which depend strongly on the vessel diameter.
Flow rates of up to \SI{500}{\cubic\cm\per\s} leave the heart and thus most of the blood flows into the lower torso ($77\%$), the head ($13 \%$) or the arms ($10 \%$). 
Only a small fraction of 1.2 per-mille, reaches the breast.
The flow through the ``anterior communicating artery'' at the very top, connecting both parts of the Circle of Willis, is very small.
This vessel typically only acts as a backup in case one of the inflows to the brain fails, and our prediction matches medical knowledge. The pressures and flows in Vessels 5 and 6, 23 and 24 as well as in 14 and 16 are nearly the same.
This is due to the high spatial symmetry of the underlying vasculature and is to be expected for a valid model.

\begin{figure} \centering
	\includegraphics[width=0.52\textwidth]{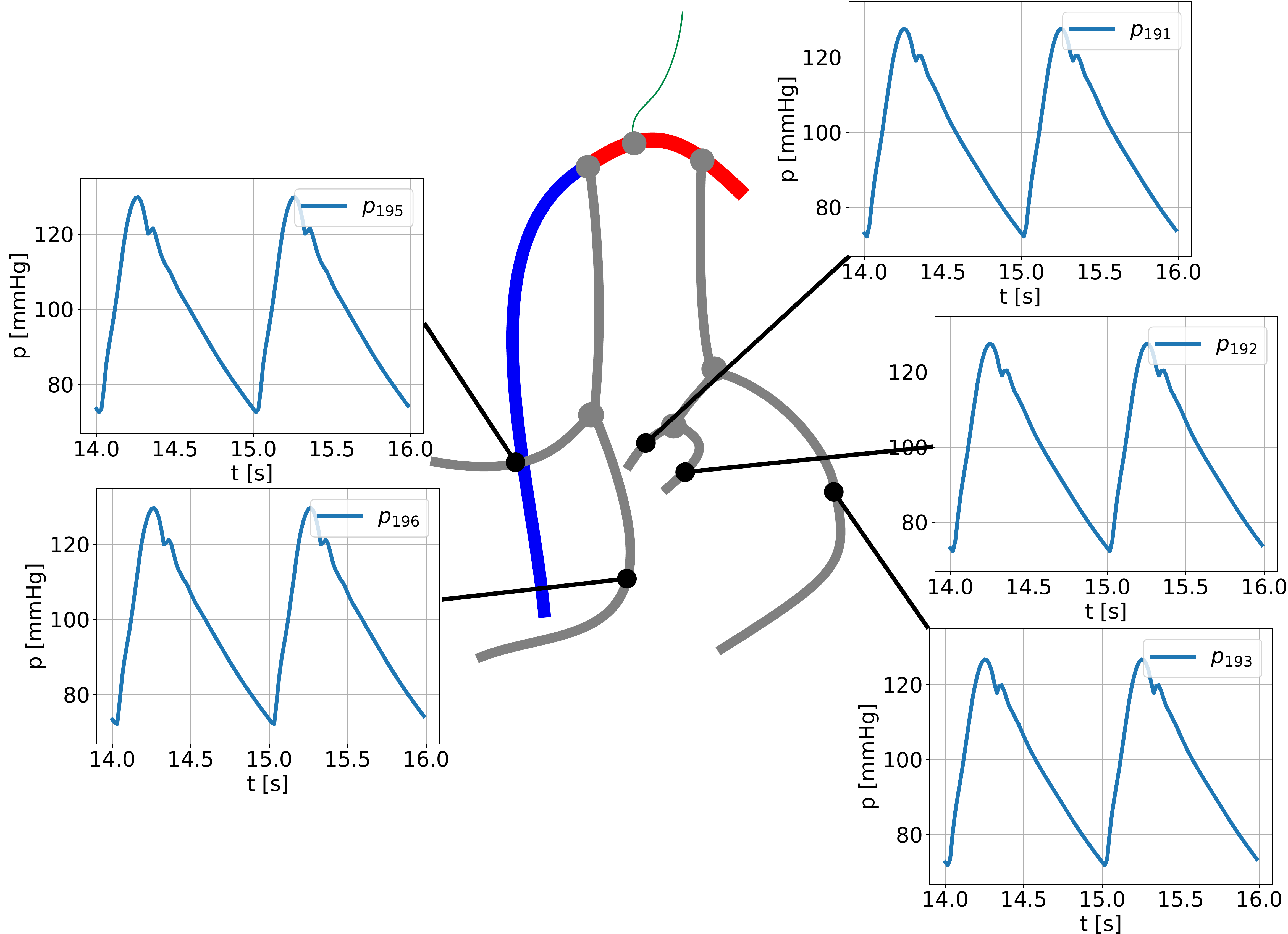}%
	\vspace{0.5cm}
	\includegraphics[width=0.52\textwidth]{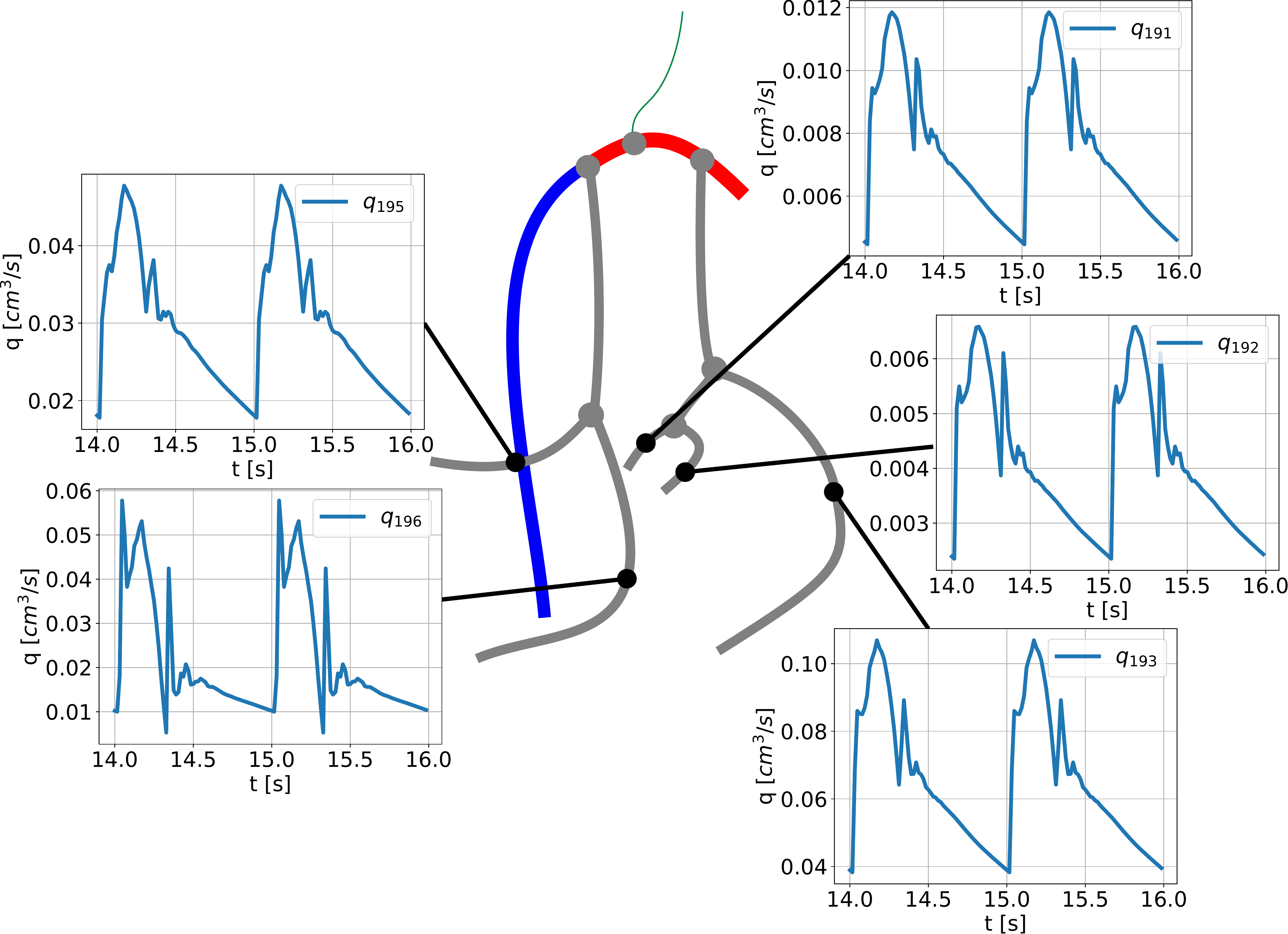}%
	\caption{\label{fig:results1DFlowsConnection}
		Pressure and flow rates within the artificial vessels connecting the large arteries with the breast geometry.}
\end{figure}

Figure~\ref{fig:results1DFlowsConnection} depicts the flow through the artificial vessels connecting the larger arteries with the measured breast geometry.
A periodic behavior is observed and flows have a similar shape. 
Compared to the flows into the brain, arms, and lower torso, the flows into the breast are small, but always positive and therefore continuously provide blood to the breast.
The high frequency components of the flow are quite possibly due to reflections at the vessel boundaries, since the arm arteries have a much larger diameter than the artificial arteries to the breast. 

\begin{figure} \centering
		\includegraphics[width=.22\textwidth]{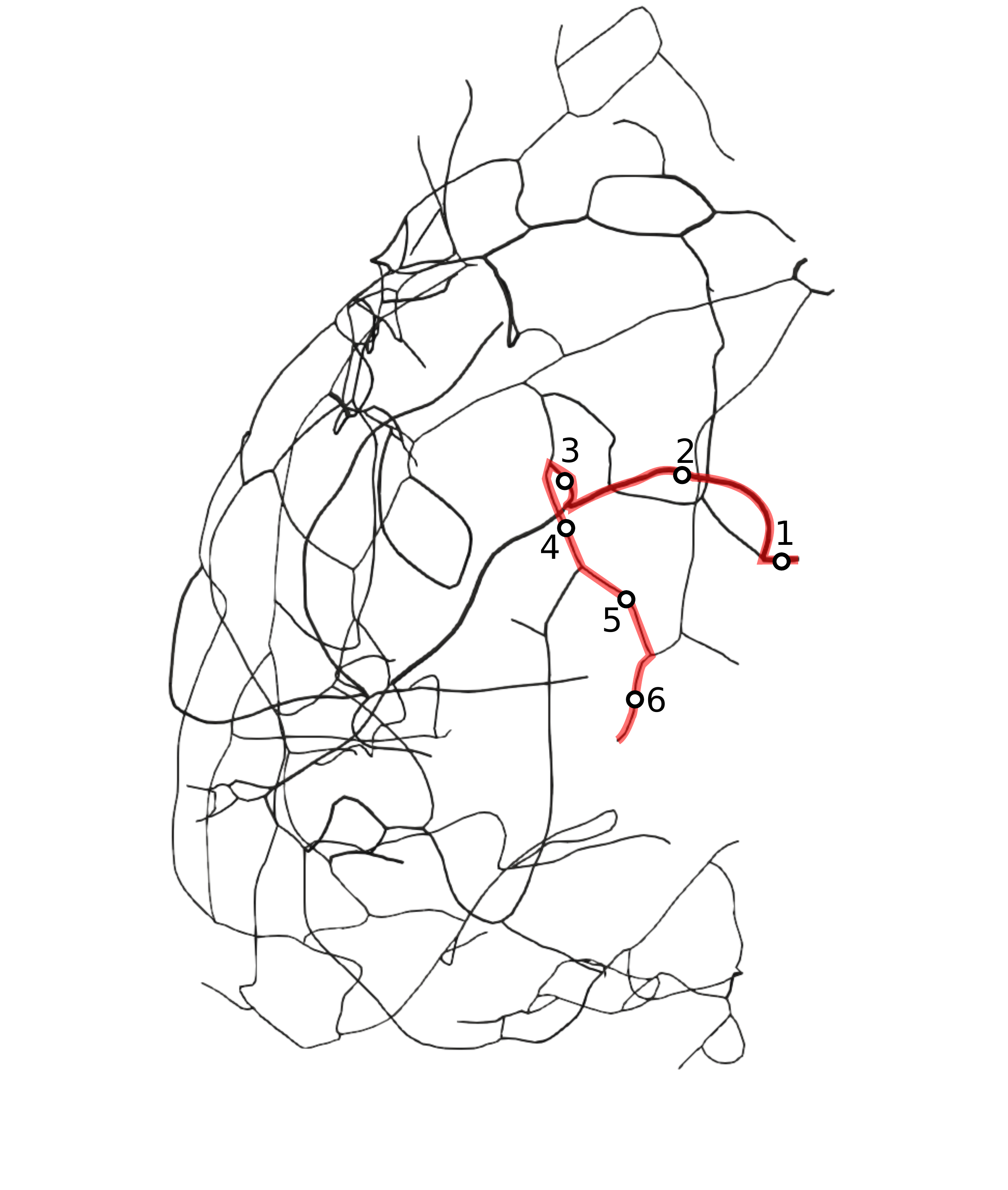}%
		\includegraphics[width=.54\textwidth]{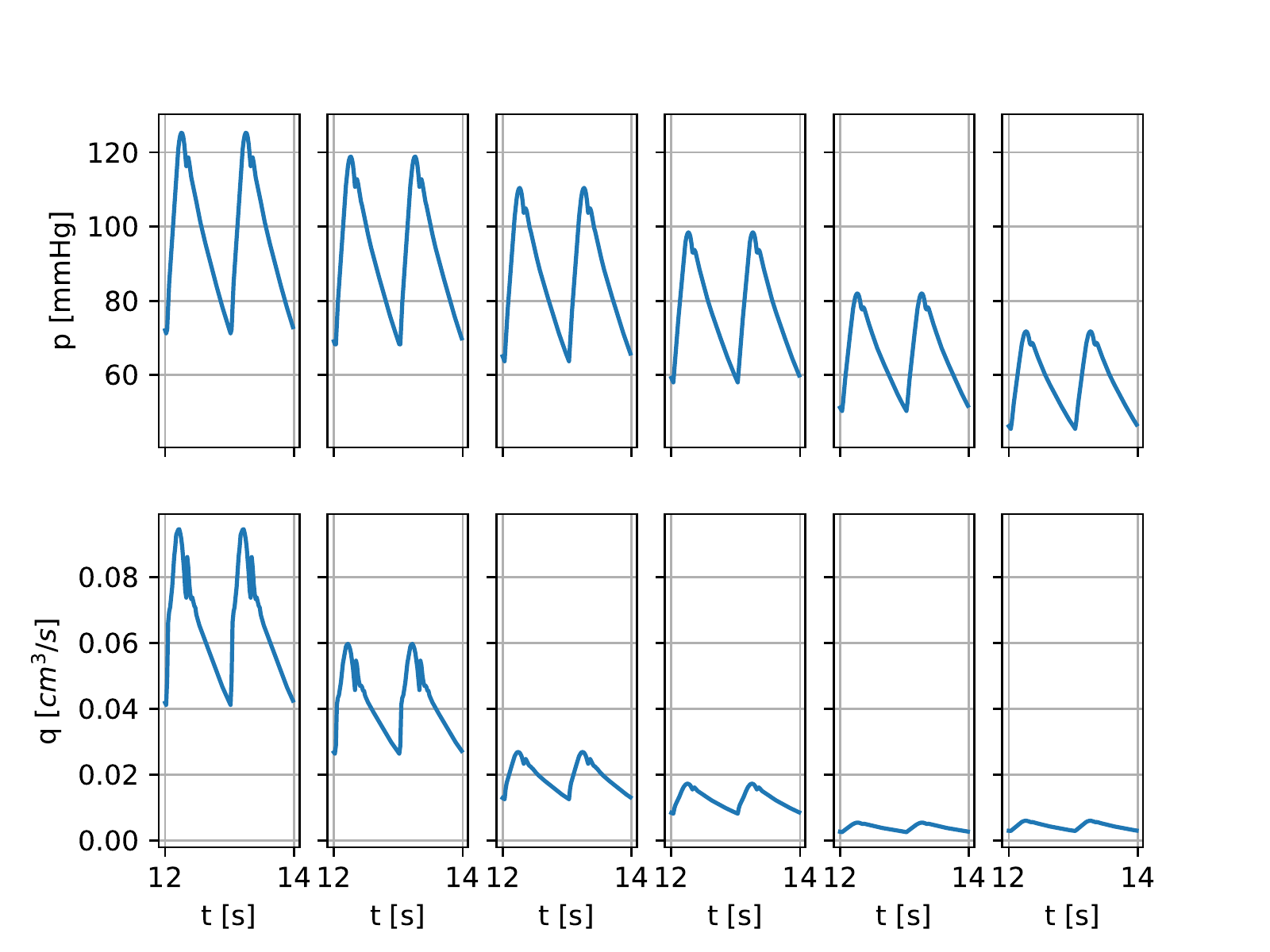}
	\caption{\label{fig:resultsLinearizedGeometry}Pressures (upper right) and flows (lower right) through one exemplary path in the breast geometry (left).  }
\end{figure}

In Figure~\ref{fig:resultsLinearizedGeometry}, the pressures and flows of the linearized model for a single flow path starting at an inlet connected to the nonlinear regime and ending at an outlet connected to our lumped parameter models are shown. It is observed that approaching the outlet, the average pressure is slowly damped down. Starting at \SI{123/75}{\mmHg}, it gets damped down to \SI{70/45}{\mmHg}.
The flow on the other hand decreases much faster, since it strongly depends on the size of the vessel which visibly decreases. Also, the high frequency components of the flow, which are visible near the coupling points and are still prominent near the inlet, get smoothed in space so that the flow reaching the outlets already appears very smooth. 

\begin{figure} \centering
	\includegraphics[width=.65\textwidth]{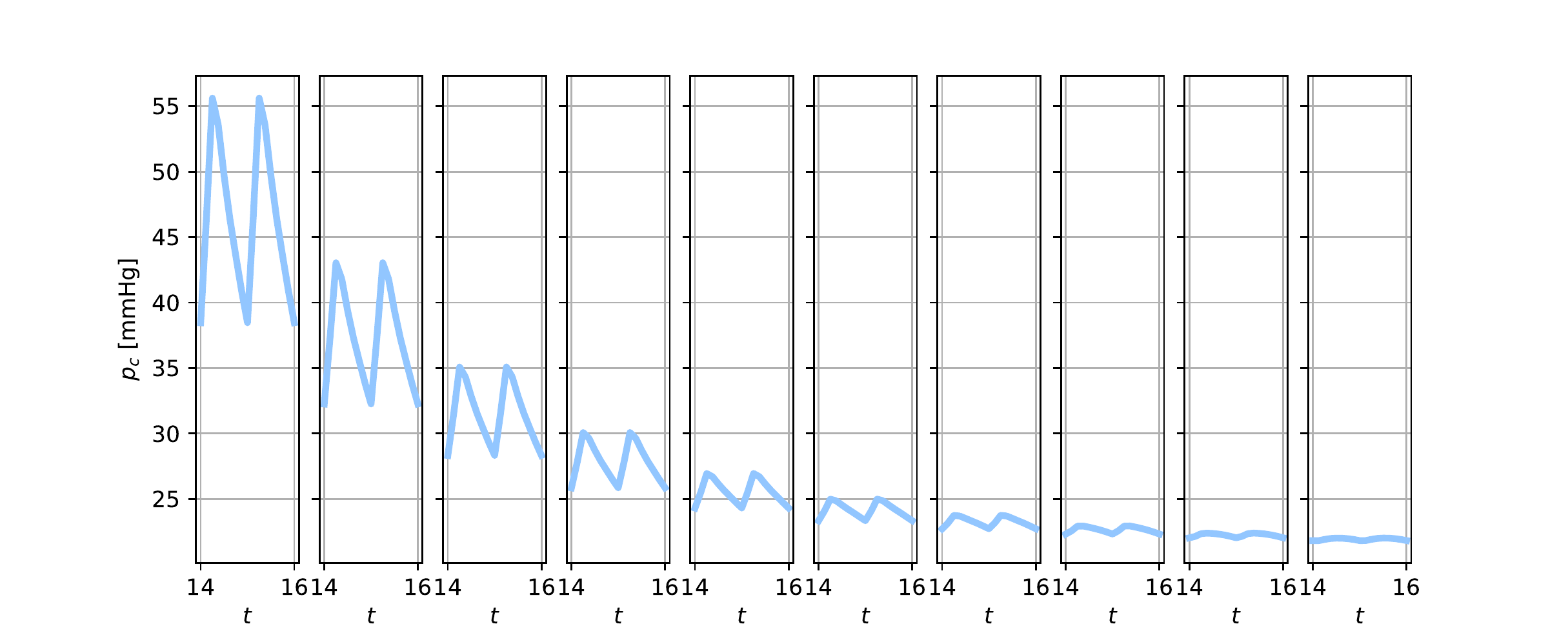}%
	\caption{\label{fig:results0DPressureSingleTree}
        Pressure decay within a 0D-tree model. For each compartment the pressure waves for two heart beats are shown.}
\end{figure}

In Figure~\ref{fig:results0DPressureSingleTree}, pressures inside the vessels of 0D-Tree models are depicted. Every subplot belongs to one lumped vessel in the hierarchy, where the leftmost vessel interacts with the linearized 1D model, while the rightmost vessel is coupled to the 3D homogenized capillary domain.
The time interval contains two heart beats from \SI{14}{\s} to \SI{16}{\s}.
After 14 heart beats, the solution is already periodic in all the compartments of our vascular tree.
In addition, a large drop in the pressure amplitude from about \SI{16}{\mmHg} to \SI{0.5}{\mmHg} is observed. The pressures in the last compartment falls into the range of capillary pressures in Figure~\ref{fig:PressureInDiffVessels}.

\begin{figure} \centering
	\includegraphics[width=.625\textwidth]{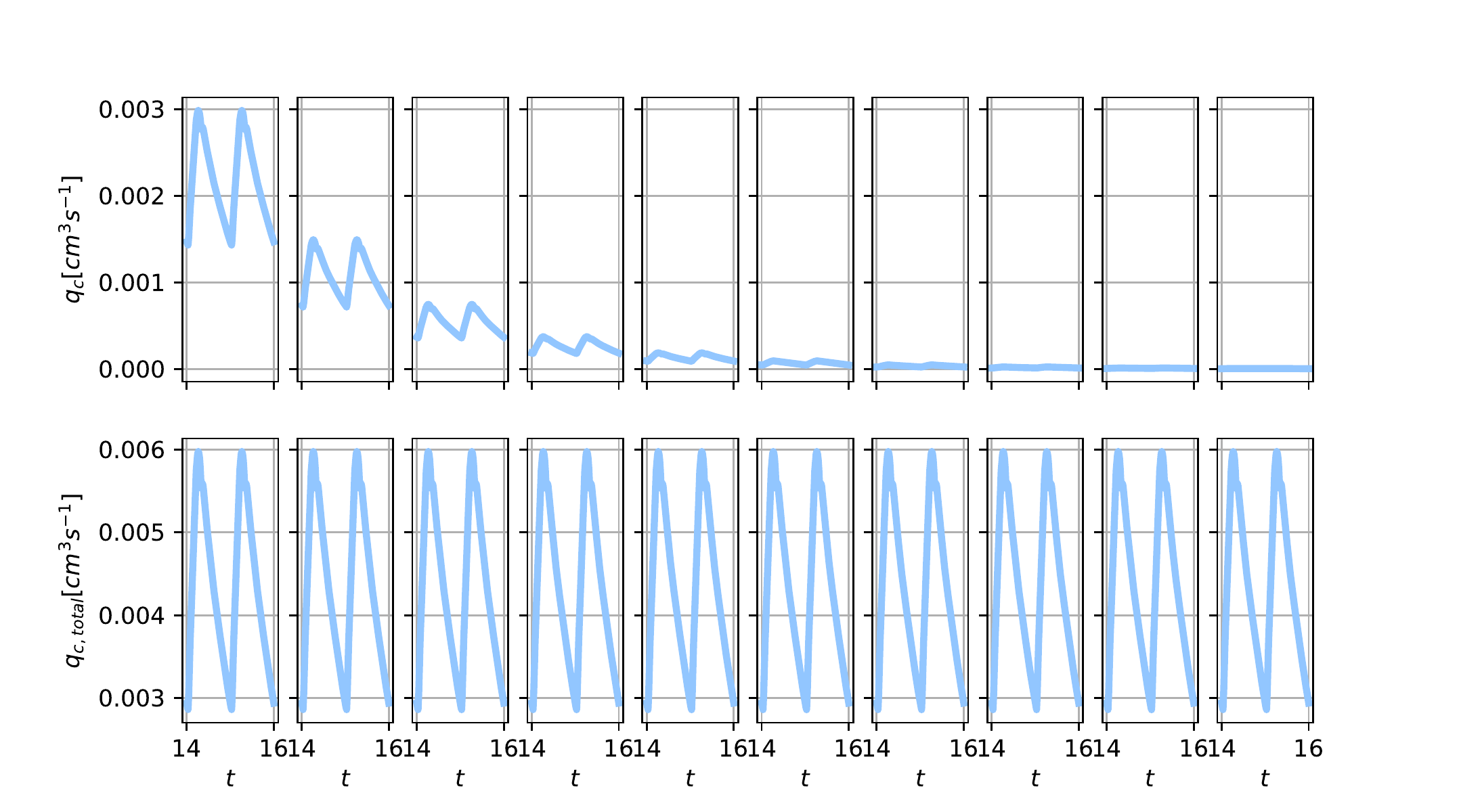}%
	\caption{\label{fig:results0DFlowsSingleTree}
        The graphs show the flow rates in one prototypical vessel of the 0D tree over time. Top: Flow rate in one vessel on each bifurcation level.
		Bottom: Total flow in all vessels of the tree belonging to one bifurcation level.}
\end{figure}

In Figure~\ref{fig:results0DFlowsSingleTree}, additional results on the flow inside the vessels of our 0D tree models are shown.
As can be seen in the upper row, for a single vessel the flow decreases by one half from one compartment level to the next due to the bifurcation.
Multiplying the flow by the number of vessels on each level replicates the total flow.
The bottom row in the figure shows that the total flow is conserved. 
Integrating the total flow of all 0D models over time yields that $\SI{1.3e-1}{\cubic\cm}$ of blood leave the 0D-networks every heart beat.
The blood volume leaving a single 0D network is between $\SI{1.3e-3}{\cubic\cm}$ and $\SI{8.3e-3}{\cubic\cm}$.

\begin{figure}\centering
	\includegraphics[width=0.35\textwidth]{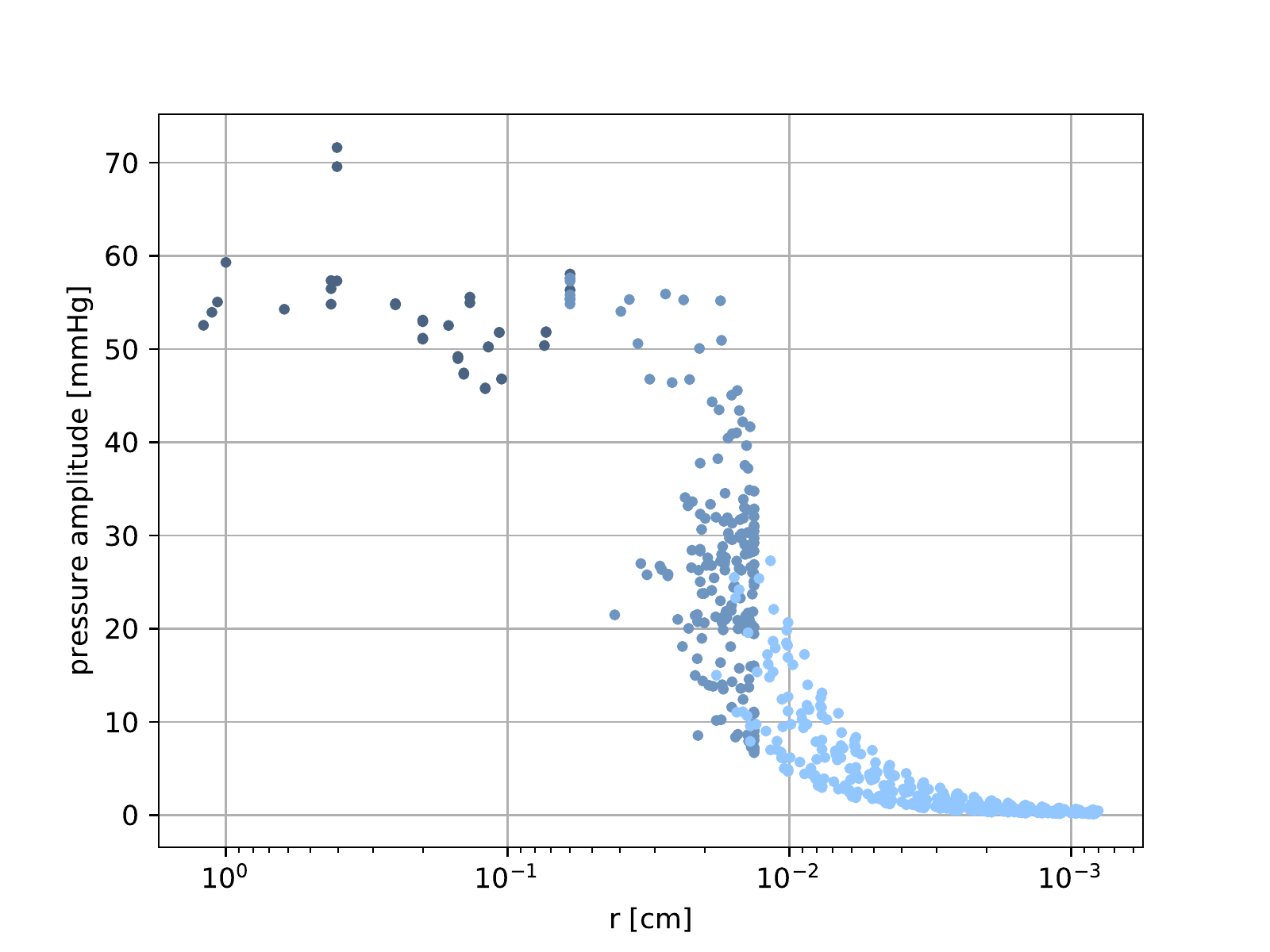}%
	\includegraphics[width=0.35\textwidth]{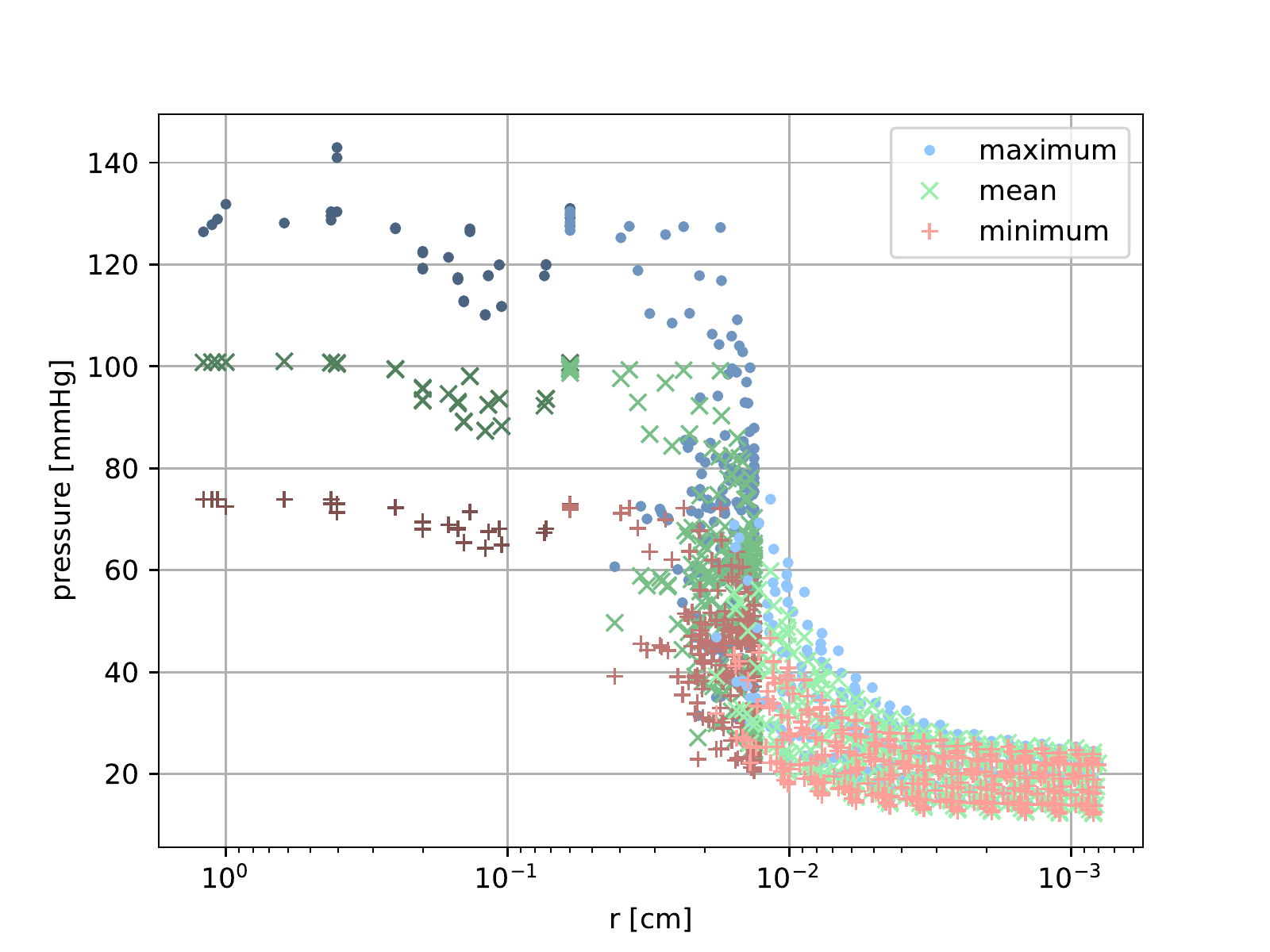}
	\caption{
		\label{fig:results1D0DAmpMaxMeanMin}
		Plot of the vessel radii of the nonlinear, linearized and 0D flow models against the pressures which were averaged from heart beat 14 to heart beat 20. 
		To differentiate between the models, the color value gets lighter from the nonlinear over the linearized to the 0D model.
		The left plot shows the pressure amplitude in this time interval, while the right shows the maximum, mean and minimum pressures for all radii.}
\end{figure}

In Figure~\ref{fig:results1D0DAmpMaxMeanMin}, the pressure amplitude on the left, as well as the mean, maximal, and minimal pressure values in our 1D and 0D models are shown.
The color gets lighter from the nonlinear, over the linearized to the 0D tree-models.
In 1D, the pressures are measured at the center of the vessels, and for the varying radii in the nonlinear regime, we use the one at zero pressure.
For large and medium-sized arteries, the pressure amplitude stays in the same range at \SI{55}{\mmHg}, and we have a stable mean pressure just below \SI{100}{\mmHg}.
Our model can replicate the effect that the pressure does not attain its maximum at the aorta, but in the brachial arteries, where we reach a pressure amplitude of \SI{72}{\mmHg}. 
For smaller arterial vessels, we see a pressure drop from \SI{50}{\mmHg} to \SI{7}{\mmHg}, which is smeared out over a large range of different pressure values. 
Here we enter the arterioles, which are modeled by means of lumped parameter models.
The pressure amplitude decreases significantly, and the mean pressure values converge for smaller radii until they are in a range between \SI{25}{\mmHg} to \SI{12}{\mmHg}.
Both plots are in agreement with the expected qualitative results of Figure~\ref{fig:PressureInDiffVessels}. It can be observed that the input pressure for the arterioles is around \SI{30}{\mmHg} and that the pressure amplitudes decrease significantly.

\begin{figure} \centering
	\includegraphics[width=.60\textwidth]{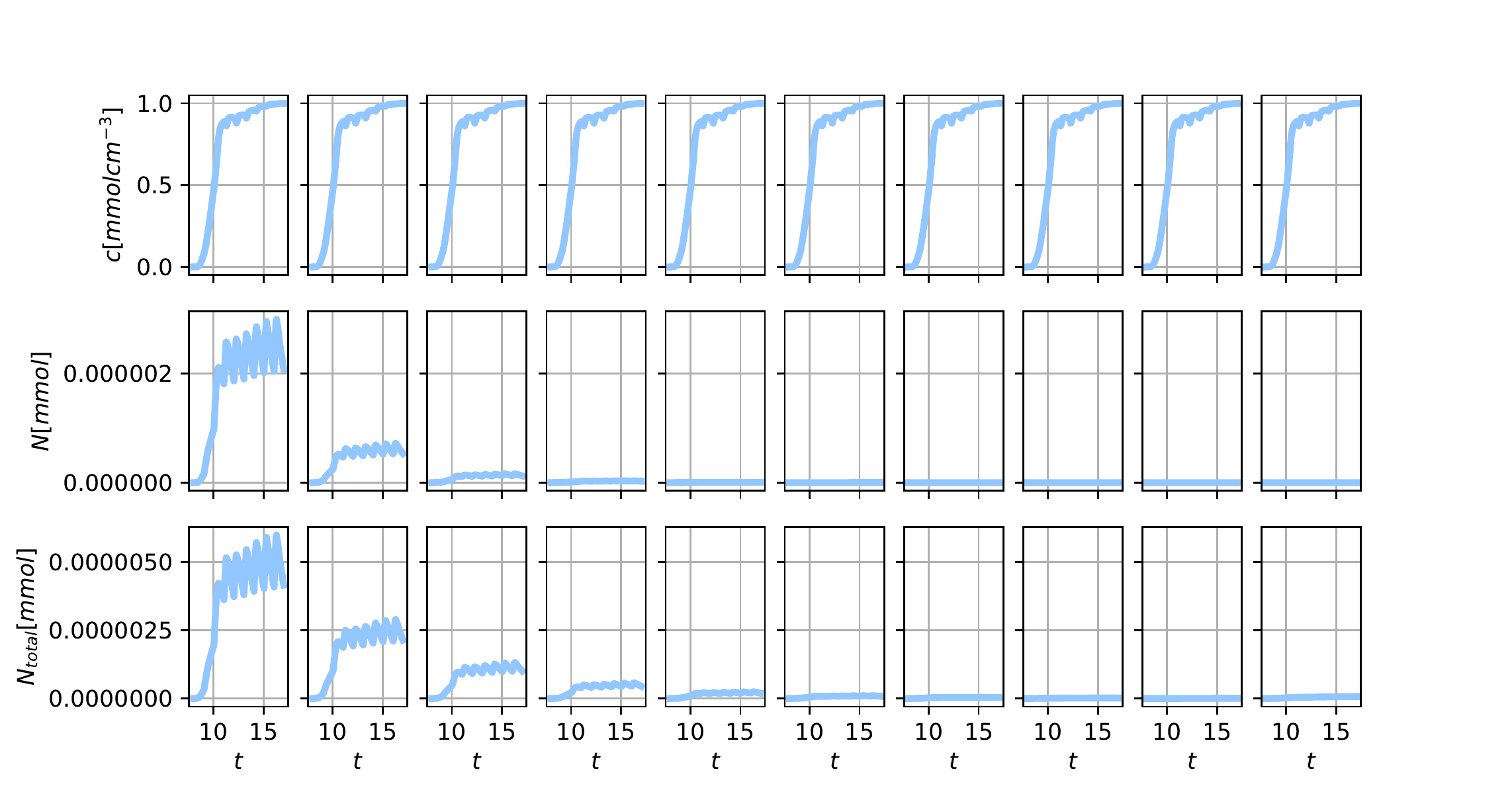}
	\caption{ \label{fig:results0DTransport} Transport in the compartments of a 0D tree. Top: Volumetric concentration for each tree level.
	Middle: The amount of substance in one vessel per level.
	Bottom: The total amount of substance in all $2^{k}$ vessels of the $k^{\textmd{th}}$ level in one tree.}
\end{figure}
To test the transport model, a time-independent source of \SI{1}{\mmol\per\cm} is placed at the aortic entrance.
The concentration front propagates through the whole network, mostly during the systolic phase of the cardiac cycle. Figure~\ref{fig:results0DTransport} depicts the transport inside the 0D trees.
The upper row depicts the concentration inside the compartments.
The concentration front enters the tree around heart beats 8 and propagates nearly instantaneously through the whole tree.
At heart beat 14 it has reached a fixed value of $1\;\left[ \unitfrac{mmol}{cm^3} \right]$.
The middle row shows the amount of substance in \SI{}{\mmol} in each compartment, while the lower row shows the total amount of substance stored in all vessels of our tree at a given hierarchy level.
Both quantities decrease from one level to the next level. Thus, only a small amount of mass is stored in the arteriol trees.
Combining these results with the flow rate of approximately \SI{2.25e-3}{\cubic\cm\per\s} at the tip of our tree from Figure~\ref{fig:results0DFlowsSingleTree}, suggests an average flow rate of \SI{2.25e-3}{\mmol\per\s} from our trees into the 3D domain.

\begin{figure} \centering
	\includegraphics[width=.18\textwidth]{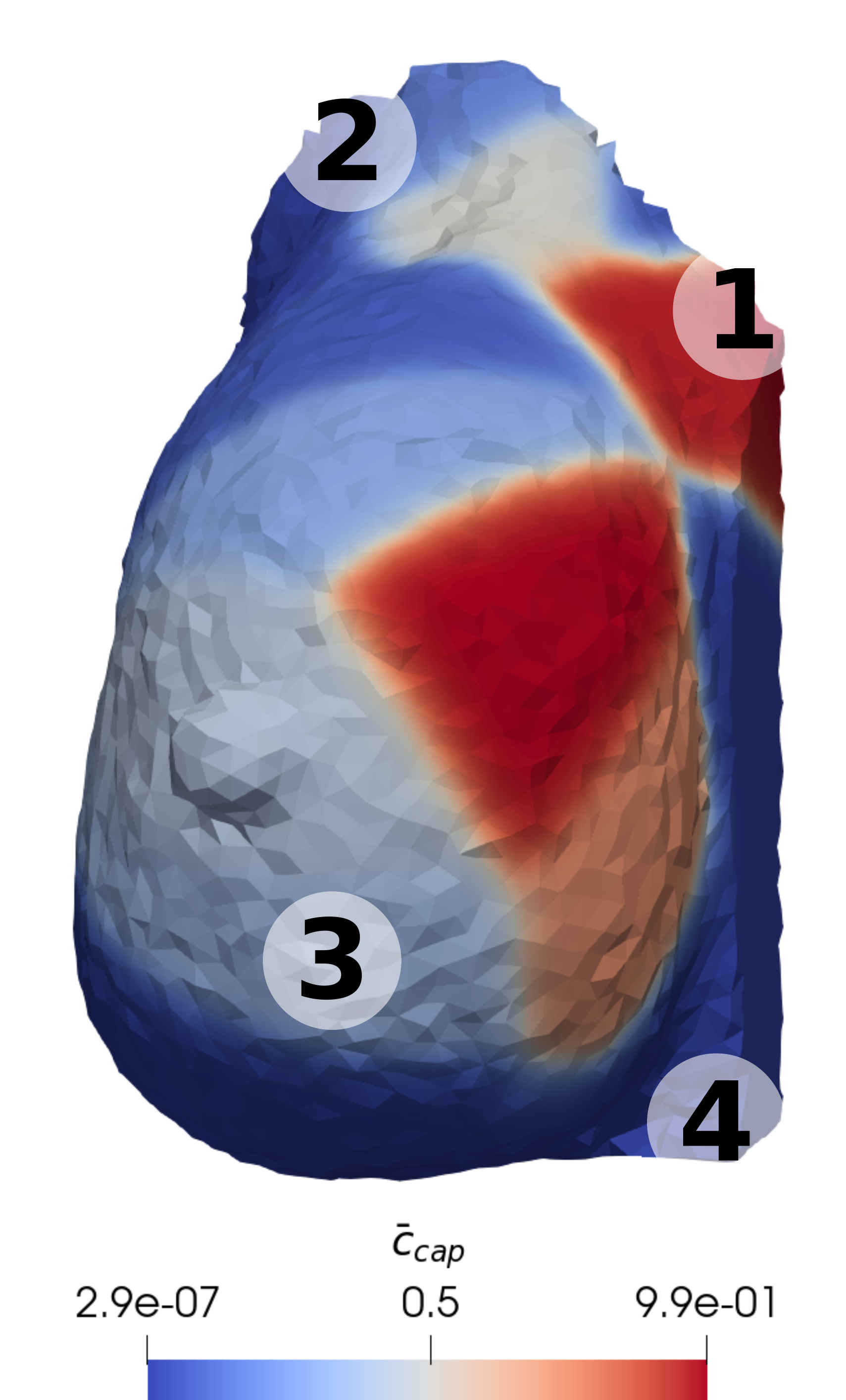} \qquad
	\includegraphics[width=.44\textwidth]{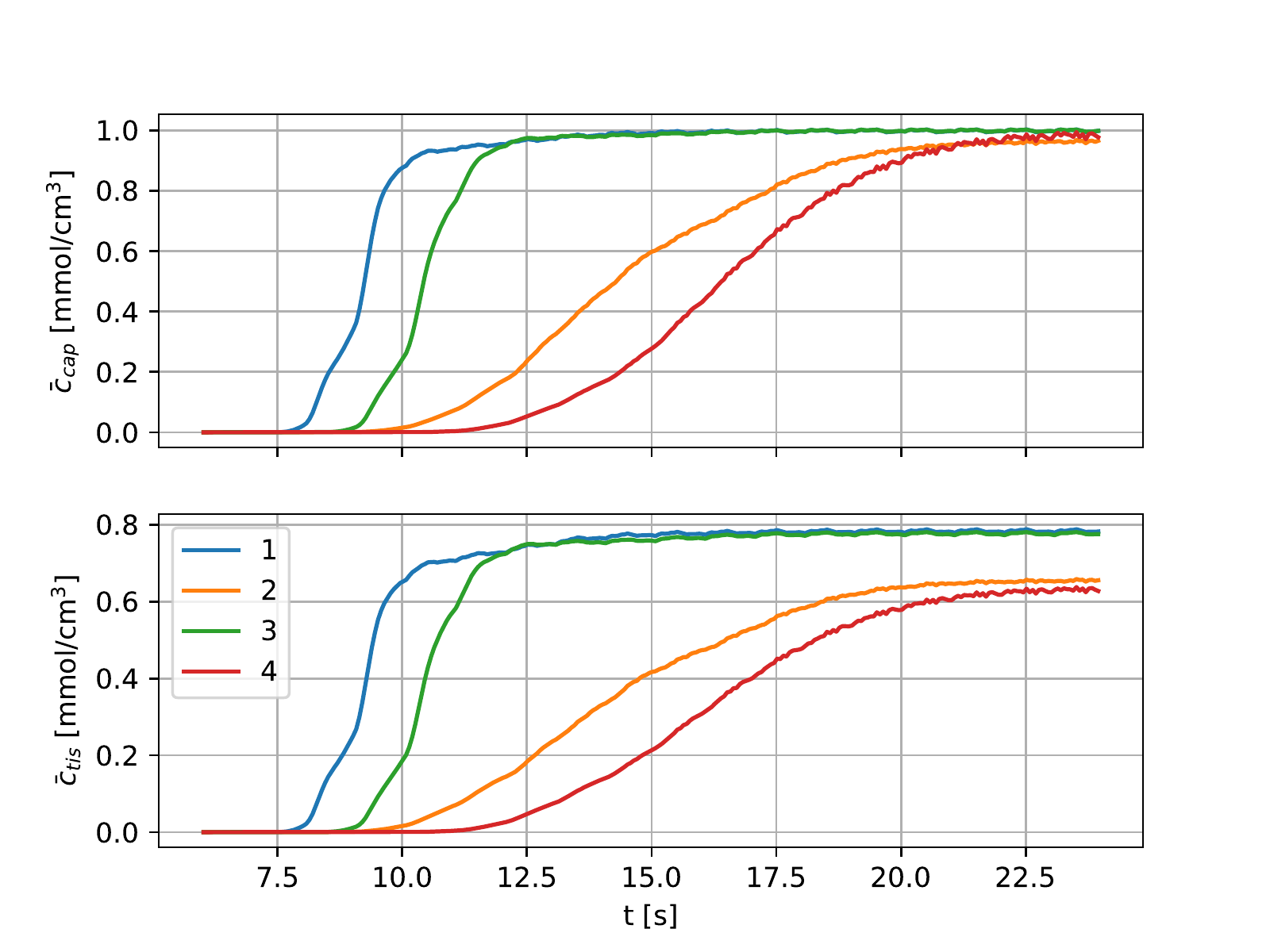}
	\caption{ \label{fig:results3DTransport}
		Left: Concentration in capillaries at $t=10$.
		Right: Average concentrations in capillaries (upper) and tissue (lower) for the four perfusion zones marked on the left.
	 }
\end{figure}
In Figure~\ref{fig:results3DTransport}, we see the concentration in capillaries and tissue of our perfusion domains. When a concentration front arrives at the vessel tips it starts to spread uniformly in its domain. The capillary concentration approaches a constant value of $\SI{1}{\milli\mole\per\cubic\cm}$, while the asymptotic mean tissue concentrations vary between $\SI{0.6}{\milli\mole\per\cubic\cm}$ and $\SI{0.8}{\milli\mole\per\cubic\cm}$.

\subsection{Breast model}
\label{sec:resultsDecoupledModel}

\begin{figure}\centering
	\includegraphics[width=.6\textwidth]{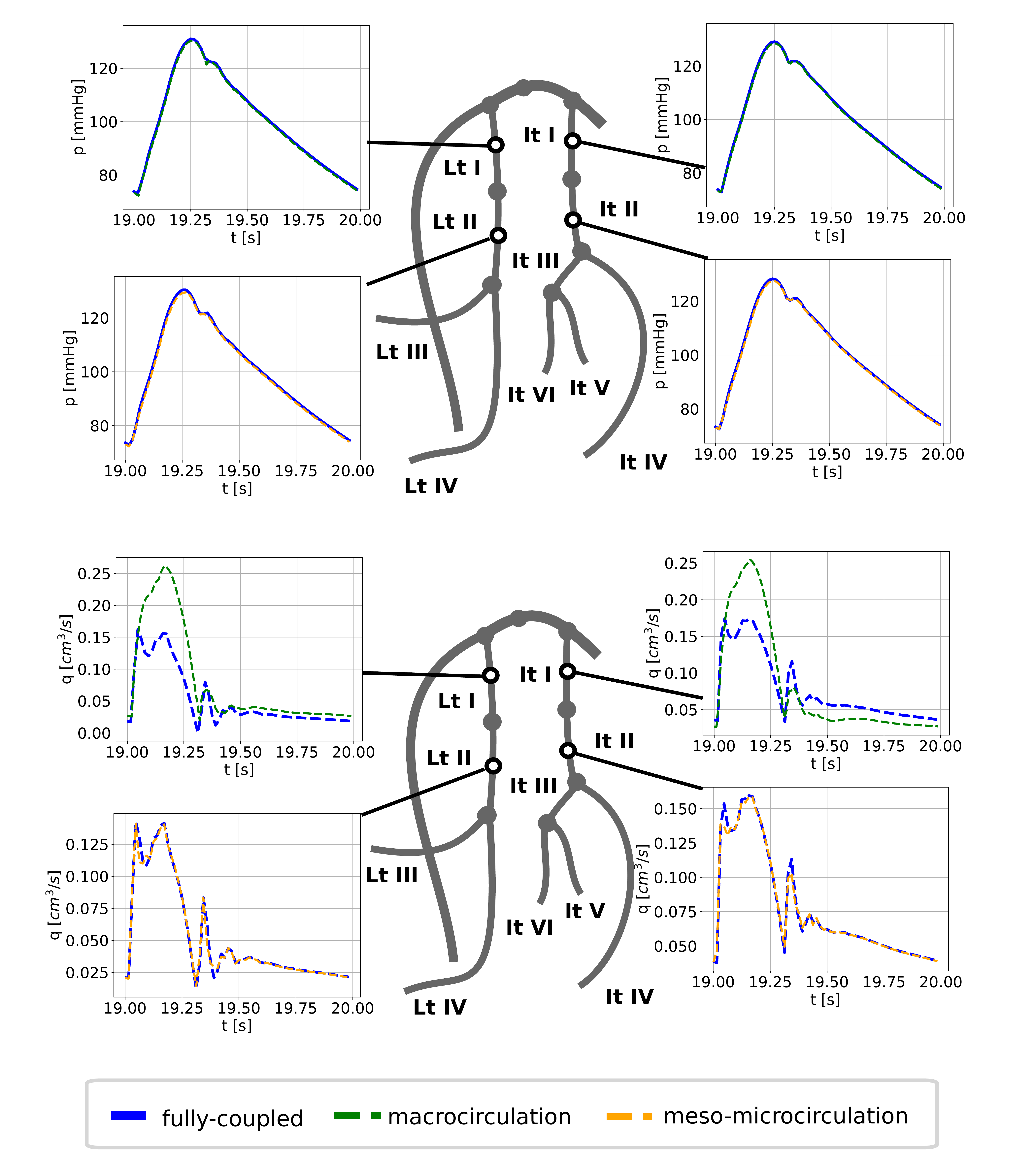}%
	\caption{ \label{fig:results0DDecoupled} Comparison between the fully coupled model and the meso-microcirculation model in the artificial extensions for the pressures and flow rates.}
\end{figure}

Next, results delivered by the much cheaper uncoupled breast model are compared with the fully coupled model. Figure~\ref{fig:results0DDecoupled} compares both models in Lt I, It I and Lt II, It II. For the pressure we observe for all four cases an excellent agreement. For the flow, the situation is significantly different. 
For Lt I and It I, the macrocirculation model with the Windkessel boundary overestimates the flow rate by $\approx 30 \si{\percent}$ and expects more fluid to enter the breast.
However for Lt II and It II, the flow rates obtained from the pressure boundary condition coincide again.

\begin{figure}\centering
	\includegraphics[width=.40\textwidth]{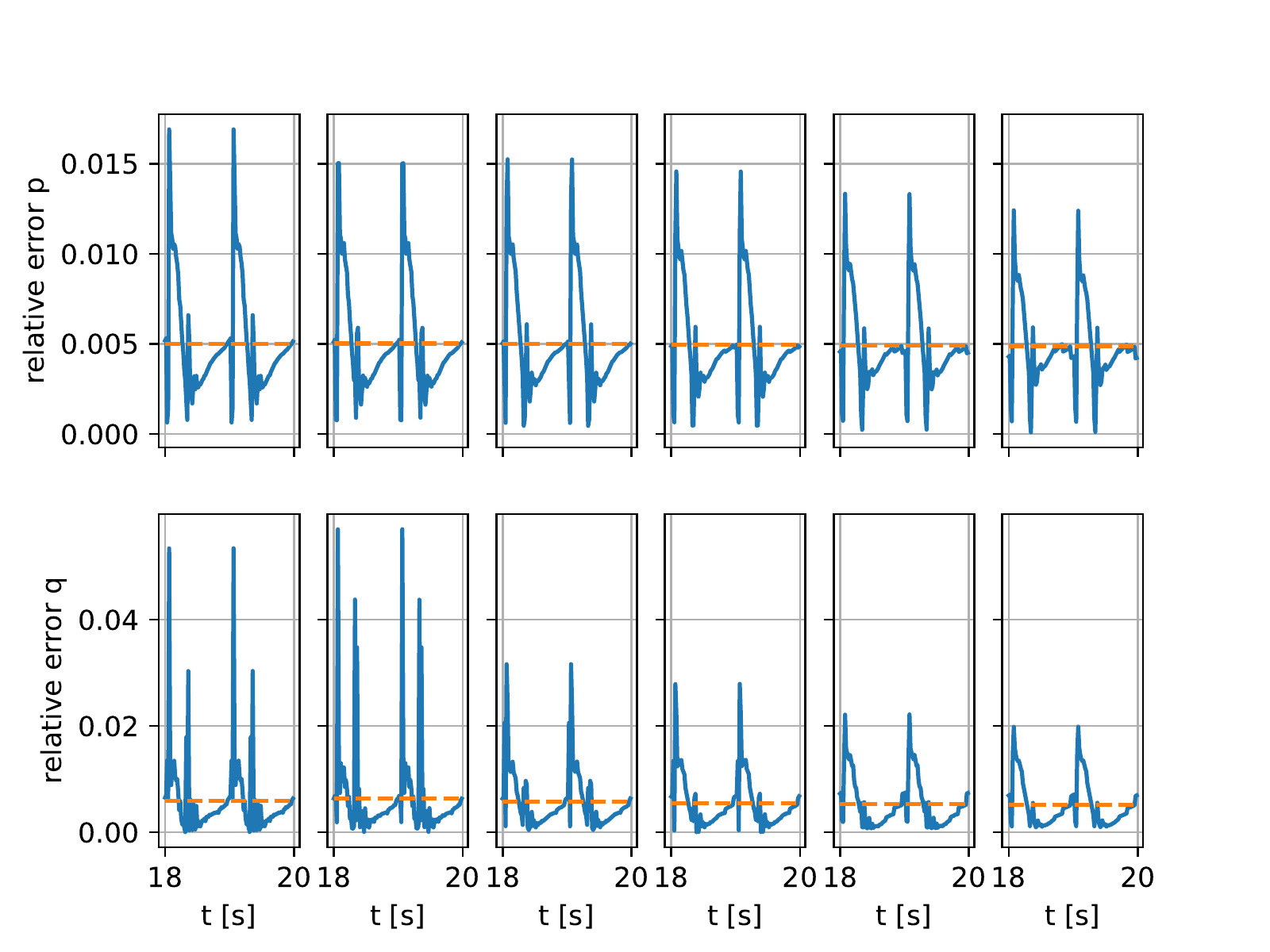}%
	\includegraphics[width=.40\textwidth]{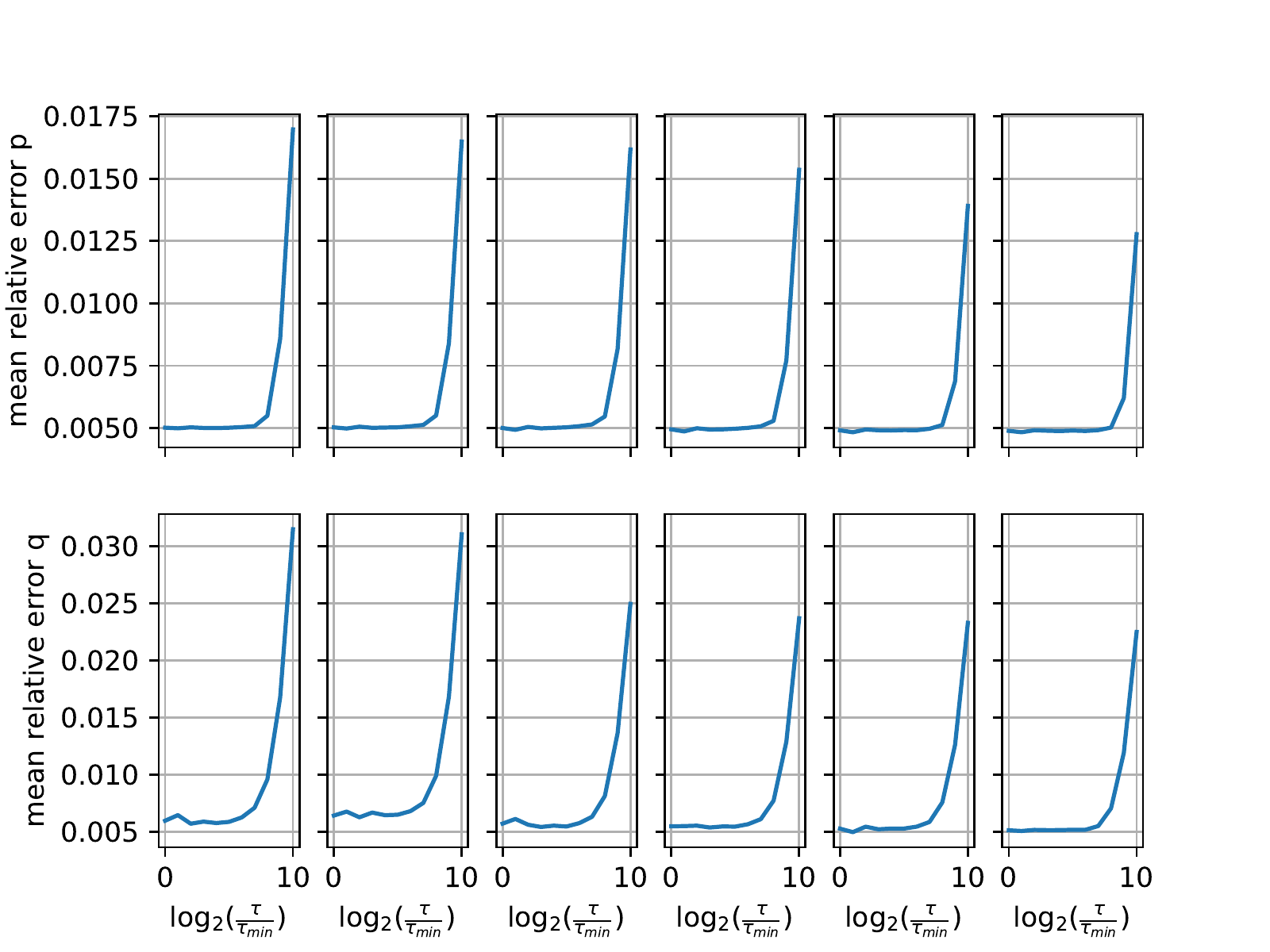}
	\caption{\label{fig:resultsLinearizedGeometryDecoupledComparison} 
	Left: Comparison between fully coupled and breast model for vessels along the path of Figure~\ref{fig:resultsLinearizedGeometry} (left). A solid line is used for the relative error, its mean in time is depicted with a dashed line.
	Right: The mean relative errors between the fully coupled model at a fixed time step $\tau_{min} = \SI{1.5625e-5}{s}$ versus the breast model with a varying time step width $\tau$. }
\end{figure}

Figure~\ref{fig:resultsLinearizedGeometryDecoupledComparison} (left) depicts the time-dependent relative error between the
fully coupled and breast model along a given path for a fixed time step width $\tau_{min}$. Its mean in time, depicted with a dashed line, is smaller than $\SI{1}{\percent}$ for both the pressure and the flow. The error itself is periodic and has peaks when the underlying pressure and flow curves change in a nonsmooth way. In space, these peaks are damped, while the mean relative error stays constant.

For the fully coupled model, the time step size cannot be increased due to the CFL condition of the explicit nonlinear parts in our 1D equations.
This is not the case for the linearized equations for which we use an implicit scheme and can choose $\tau$ without any restriction.
Figure~\ref{fig:resultsLinearizedGeometryDecoupledComparison} (right) depicts the mean error between the fully coupled model at a fixed time step width $\tau_{min}$ versus the breast model with varying $\tau$.
For a large range of $\tau$, the modeling error dominates. Thus, even increasing $\tau$ by a factor of $100$ does not change the relative error. For larger $\tau$ the error increases first in the vessels near the inlets.
If we are only interested in the flow at the outlets, this justifies even more the usage of larger time step sizes.

%% file: multiscale_8_conclusions.tex
\section{Conclusion and outlook}
\label{sec:conclusion}

A class of multiscale models has been introduced in this investigation to simulate flow and transport from heart to breast. 
To simulate the pressure amplitudes in a realistic way, a nonlinear model was used for the arteries, a linearized model for the smaller arteries, and a 0D model for the arterioles. Then this model has been coupled to the interstitial space via the homogenized 3D capillary model. Our findings are qualitatively consistent with medical knowledge about the circulation of blood within the human body. A fully coupled model has been described and compared to a simpler sub-model. It has been demonstrated experimentally that they produce sufficiently similar results for all practical purposes. The breast model consists of a nonlinear model on the patient independent parts and a linearized model for the patient specific parts. This suggests, that the expensive nonlinear computation can be done once for a large class of patients in an offline computation. For the patient specific parts, we can use use a fully implicit solver allowing us to cheaply bridge long simulation times by using large time step sizes. 

Results also suggest that precompiling the outlet pressures with the macrocirculation model and feeding these into the breast networks of various patients is sufficiently accurate for use in future studies of drug delivery and tumor growth.

%% file: paper_multiscale_bloodflow.bbl
\begin{thebibliography}{10}

\bibitem{alastruey2007modelling}
J.~Alastruey, K.~Parker, J.~Peir{\'o}, S.~Byrd, and S.~Sherwin.
\newblock {Modelling the Circle of Willis to assess the effects of anatomical
  variations and occlusions on cerebral flows}.
\newblock {\em Journal of Biomechanics}, 40(8):1794--1805, 2007.

\bibitem{alastruey2008lumped}
J.~Alastruey, K.~Parker, J.~Peir{\'o}, and S.~Sherwin.
\newblock {Lumped parameter outflow models for 1-D blood flow simulations:
  Effect on pulse waves and parameter estimation}.
\newblock {\em Communications in Computational Physics}, 4(2):317--336, 2008.

\bibitem{petsc-user-ref}
S.~Balay, S.~Abhyankar, M.~F. Adams, S.~Benson, J.~Brown, et~al.
\newblock {PETSc/TAO Users Manual}.
\newblock Technical Report ANL-21/39 - Revision 3.16, Argonne National
  Laboratory, 2021.

\bibitem{barral2011visceral}
J.-P. Barral and A.~Croibier.
\newblock {\em {Visceral Vascular Manipulations}}.
\newblock Churchill Livingstone, 2012.

\bibitem{biga2020anatomy}
L.~M. Biga, S.~Dawson, A.~Harwell, R.~Hopkins, J.~Kaufmann, et~al.
\newblock {\em Anatomy \& Physiology}.
\newblock OpenStax/Oregon State University, 2019.

\bibitem{blanco2009potentialities}
P.~J. Blanco, M.~Pivello, S.~Urquiza, and R.~Feij{\'o}o.
\newblock On the potentialities of 3d--1d coupled models in hemodynamics
  simulations.
\newblock {\em Journal of biomechanics}, 42(7):919--930, 2009.

\bibitem{butros2014direct}
S.~R. Butros, T.~G. Walker, G.~M. Salazar, S.~P. Kalva, R.~Oklu, S.~Wicky, and
  S.~Ganguli.
\newblock Direct translumbar inferior vena cava ports for long-term central
  venous access in patients with cancer.
\newblock {\em Journal of Vascular and Interventional Radiology},
  25(4):556--560, 2014.

\bibitem{vcanic2003mathematical}
S.~{\v{C}}ani{\'c} and E.~Kim.
\newblock {Mathematical analysis of the quasilinear effects in a hyperbolic
  model blood flow through compliant axi-symmetric vessels}.
\newblock {\em Mathematical Methods in the Applied Sciences},
  26(14):1161--1186, 2003.

\bibitem{cattaneo2014fem}
L.~Cattaneo.
\newblock {\em {FEM for PDEs with Unfitted Interfaces: Application to Flow
  through Heterogeneous Media and Microcirculation}}.
\newblock PhD thesis, Polytechnic University of Milan, 2014.

\bibitem{d2007multiscale}
C.~D'Angelo.
\newblock {\em {Multiscale Modelling of Metabolism and Transport Phenomena in
  Living Tissues}}.
\newblock PhD thesis, \'Ecole Polytechnique F\'ed\'erale de Lausanne, 2007.

\bibitem{dewhirst2017transport}
M.~Dewhirst and T.~Secomb.
\newblock {Transport of drugs from blood vessels to tumour tissue}.
\newblock {\em Nature Reviews Cancer}, 17(12):738--750, 2017.

\bibitem{di2021computational}
S.~Di~Gregorio, M.~Fedele, G.~Pontone, A.~Corno, P.~Zunino, C.~Vergara, and
  A.~Quarteroni.
\newblock A computational model applied to myocardial perfusion in the human
  heart: from large coronaries to microvasculature.
\newblock {\em Journal of Computational Physics}, 424:109836, 2021.

\bibitem{drzisga2016numerical}
D.~Drzisga, T.~K{\"o}ppl, U.~Pohl, R.~Helmig, and B.~Wohlmuth.
\newblock {Numerical modeling of compensation mechanisms for peripheral
  arterial stenoses}.
\newblock {\em Computers in Biology and Medicine}, 70:190--201, 2016.

\bibitem{ehlers2015multi}
W.~Ehlers and A.~Wagner.
\newblock {Multi-component modelling of human brain tissue: A contribution to
  the constitutive and computational description of deformation, flow and
  diffusion processes with application to the invasive drug-delivery problem}.
\newblock {\em Computer methods in Biomechanics and Biomedical Engineering},
  18(8):861--879, 2015.

\bibitem{el2018investigating}
W.~El-Bouri and S.~Payne.
\newblock {Investigating the effects of a penetrating vessel occlusion with a
  multi-scale microvasculature model of the human cerebral cortex}.
\newblock {\em NeuroImage}, 172:94--106, 2018.

\bibitem{erbertseder2012coupled}
K.~Erbertseder, J.~Reichold, B.~Flemisch, P.~Jenny, and R.~Helmig.
\newblock {A coupled discrete/continuum model for describing cancer-therapeutic
  transport in the lung}.
\newblock {\em PLoS One}, 7(3):e31966, 2012.

\bibitem{fernandez2005analysis}
M.~Fern{\'a}ndez, V.~Milisic, and A.~Quarteroni.
\newblock {Analysis of a geometrical multiscale blood flow model based on the
  coupling of ODEs and hyperbolic PDEs}.
\newblock {\em Multiscale Modeling \& Simulation}, 4(1):215--236, 2005.

\bibitem{formaggia2001coupling}
L.~Formaggia, J.-F. Gerbeau, F.~Nobile, and A.~Quarteroni.
\newblock {On the coupling of 3D and 1D Navier--Stokes equations for flow
  problems in compliant vessels}.
\newblock {\em Computer Methods in Applied Mechanics and Engineering},
  191(6-7):561--582, 2001.

\bibitem{formaggia2003one}
L.~Formaggia, D.~Lamponi, and A.~Quarteroni.
\newblock {One-dimensional models for blood flow in arteries}.
\newblock {\em Journal of engineering mathematics}, 47(3):251--276, 2003.

\bibitem{formaggia2002one}
L.~Formaggia, F.~Nobile, and A.~Quarteroni.
\newblock {A one dimensional model for blood flow: Application to vascular
  prosthesis}.
\newblock In {\em Mathematical Modeling and Numerical Simulation in Continuum
  Mechanics}, pages 137--153. Springer, 2002.

\bibitem{Quarteroni}
L.~Formaggia, A.~Quarteroni, and A.~Veneziani.
\newblock {\em {Cardiovascluar Mathematics: Modelling and Simulation of the
  Circulatory System}}.
\newblock Springer, 2009.

\bibitem{fritz2021modeling}
M.~Fritz, P.~K. Jha, T.~K{\"o}ppl, J.~T. Oden, A.~Wagner, et~al.
\newblock {Modeling and simulation of vascular tumors embedded in evolving
  capillary networks}.
\newblock {\em Computer Methods in Applied Mechanics and Engineering},
  384:113975, 2021.

\bibitem{eigenweb}
G.~Guennebaud, B.~Jacob, et~al.
\newblock {Eigen v3}.
\newblock http://eigen.tuxfamily.org, 2010.

\bibitem{ho2009hybrid}
H.~Ho, G.~Sands, H.~Schmid, K.~Mithraratne, G.~Mallinson, and P.~Hunter.
\newblock A hybrid 1d and 3d approach to hemodynamics modelling for a
  patient-specific cerebral vasculature and aneurysm.
\newblock In {\em International Conference on Medical Image Computing and
  Computer-Assisted Intervention}, pages 323--330. Springer, 2009.

\bibitem{hodneland2019new}
E.~Hodneland, E.~Hanson, O.~S{\ae}vareid, G.~N{\ae}vdal, A.~Lundervold, et~al.
\newblock {A new framework for assessing subject-specific whole brain
  circulation and perfusion using MRI-based measurements and a multi-scale
  continuous flow model}.
\newblock {\em PLoS Computational Biology}, 15(6):e1007073, 2019.

\bibitem{hughes1973one}
T.~Hughes and J.~Lubliner.
\newblock {On the one-dimensional theory of blood flow in the larger vessels}.
\newblock {\em Mathematical Biosciences}, 18(1):161--170, 1973.

\bibitem{jonavsova2014complex}
A.~Jon{\'a}{\v{s}}ov{\'a}, E.~Rohan, V.~Luke{\v{s}}, and O.~Bubl{\i}k.
\newblock {Complex hierarchical modeling of the dynamic perfusion test:
  application to liver}.
\newblock {\em Proceedings of the 11th World Congress on Computational
  Mechanics (WCCM2014), Barcelona}, page 2911, 2014.

\bibitem{jozsa2021porous}
T.~I. J{\'o}zsa, R.~M. Padmos, N.~Samuels, W.~El-Bouri, A.~G. Hoekstra, et~al.
\newblock {A porous circulation model of the human brain for in silico clinical
  trials in ischaemic stroke}.
\newblock {\em Interface Focus}, 11(1):20190127, 2021.

\bibitem{khaled2003role}
A.~Khaled and K.~Vafai.
\newblock {The role of porous media in modeling flow and heat transfer in
  biological tissues}.
\newblock {\em International Journal of Heat and Mass Transfer},
  46(26):4989--5003, 2003.

\bibitem{koch2020modeling}
T.~Koch, M.~Schneider, R.~Helmig, and P.~Jenny.
\newblock {Modeling tissue perfusion in terms of 1d-3d embedded mixed-dimension
  coupled problems with distributed sources}.
\newblock {\em Journal of Computational Physics}, 410:109370, 2020.

\bibitem{kojic2017composite}
M.~Kojic, M.~Milosevic, V.~Simic, E.~Koay, J.~Fleming, et~al.
\newblock {A composite smeared finite element for mass transport in capillary
  systems and biological tissue}.
\newblock {\em Computer Methods in Applied Mechanics and Engineering},
  324:413--437, 2017.

\bibitem{koeppl2018numerical}
T.~K\"oppl, G.~Santin, B.~Haasdonk, and R.~Helmig.
\newblock {Numerical modelling of a peripheral arterial stenosis using
  dimensionally reduced models and kernel methods}.
\newblock {\em International Journal for Numerical Methods in Biomedical
  Engineering}, 34(8):e3095, 2018.

\bibitem{koppl20203d}
T.~K{\"o}ppl, E.~Vidotto, and B.~Wohlmuth.
\newblock {A 3D-1D coupled blood flow and oxygen transport model to generate
  microvascular networks}.
\newblock {\em International Journal for Numerical Methods in Biomedical
  Engineering}, 36(10):e3386, 2020.

\bibitem{koppl2013reduced}
T.~K{\"o}ppl, B.~Wohlmuth, and R.~Helmig.
\newblock {Reduced one-dimensional modelling and numerical simulation for mass
  transport in fluids}.
\newblock {\em International Journal for Numerical Methods in Fluids},
  72(2):135--156, 2013.

\bibitem{kremheller2021validation}
J.~Kremheller, S.~Brandstaeter, B.~A. Schrefler, and W.~A. Wall.
\newblock Validation and parameter optimization of a hybrid
  embedded/homogenized solid tumor perfusion model.
\newblock {\em International Journal for Numerical Methods in Biomedical
  Engineering}, 37(8):e3508, 2021.

\bibitem{kremheller2019approach}
J.~Kremheller, A.-T. Vuong, B.~A. Schrefler, and W.~A. Wall.
\newblock {An approach for vascular tumor growth based on a hybrid
  embedded/homogenized treatment of the vasculature within a multiphase porous
  medium model}.
\newblock {\em International Journal for Numerical Methods in Biomedical
  Engineering}, 35(11):e3253, 2019.

\bibitem{kuzmin2010vertex}
D.~Kuzmin.
\newblock A vertex-based hierarchical slope limiter for p-adaptive
  discontinuous galerkin methods.
\newblock {\em Journal of computational and applied mathematics},
  233(12):3077--3085, 2010.

\bibitem{kuzmin2014hierarchical}
D.~Kuzmin.
\newblock Hierarchical slope limiting in explicit and implicit discontinuous
  galerkin methods.
\newblock {\em Journal of Computational Physics}, 257:1140--1162, 2014.

\bibitem{levick2010Microvascular}
J.~R. Levick and C.~C. Michel.
\newblock {Microvascular fluid exchange and the revised Starling principle}.
\newblock {\em Cardiovascular Research}, 87(2):198--210, 2010.

\bibitem{masri2021discontinuous}
R.~Masri, C.~Puelz, and B.~Riviere.
\newblock {A discontinuous Galerkin method for blood flow and solute transport
  in one-dimensional vessel networks}.
\newblock {\em Communications on Applied Mathematics and Computation}, pages
  1--30, 2021.

\bibitem{masri2021reduced}
R.~Masri, C.~Puelz, and B.~Riviere.
\newblock {A reduced model for solute transport in compliant blood vessels with
  arbitrary axial velocity profile}.
\newblock {\em International Journal of Heat and Mass Transfer}, 176:121379,
  2021.

\bibitem{mcguire2003estimation}
B.~J. McGuire and T.~Secomb.
\newblock Estimation of capillary density in human skeletal muscle based on
  maximal oxygen consumption rates.
\newblock {\em American Journal of Physiology-Heart and Circulatory
  Physiology}, 285(6):H2382--H2391, 2003.

\bibitem{murray1926physiological2}
C.~Murray.
\newblock {The physiological principle of minimum work applied to the angle of
  branching of arteries}.
\newblock {\em The Journal of General Physiology}, 9(6):835--841, 1926.

\bibitem{murray1926physiological}
C.~Murray.
\newblock {The physiological principle of minimum work: I. The vascular system
  and the cost of blood volume}.
\newblock {\em Proceedings of the National Academy of Sciences of the United
  States of America}, 12(3):207--214, 1926.

\bibitem{olufsen1999structured}
M.~S. Olufsen.
\newblock Structured tree outflow condition for blood flow in larger systemic
  arteries.
\newblock {\em American journal of physiology-Heart and circulatory
  physiology}, 276(1):H257--H268, 1999.

\bibitem{padmos2021coupling}
R.~M. Padmos, T.~I. J{\'o}zsa, W.~K. El-Bouri, P.~R. Konduri, S.~J. Payne,
  et~al.
\newblock {Coupling one-dimensional arterial blood flow to three-dimensional
  tissue perfusion models for in silico trials of acute ischaemic stroke}.
\newblock {\em Interface Focus}, 11(1):20190125, 2021.

\bibitem{peyrounette2018multiscale}
M.~Peyrounette, Y.~Davit, M.~Quintard, and S.~Lorthois.
\newblock {Multiscale modelling of blood flow in cerebral microcirculation:
  Details at capillary scale control accuracy at the level of the cortex}.
\newblock {\em PloS one}, 13(1):e0189474, 2018.

\bibitem{possenti2019computational}
L.~Possenti, S.~di~Gregorio, F.~Gerosa, G.~Raimondi, G.~Casagrande, et~al.
\newblock {A computational model for microcirculation including
  Fahraeus--Lindqvist effect, plasma skimming and fluid exchange with the
  tissue interstitium}.
\newblock {\em International Journal for Numerical Methods in Biomedical
  Engineering}, 35(3):e3165, 2019.

\bibitem{pries1996biophysical}
A.~Pries, T.~Secomb, and P.~Gaehtgens.
\newblock {Biophysical aspects of blood flow in the microvasculature}.
\newblock {\em Cardiovascular Research}, 32(4):654--667, 1996.

\bibitem{puel1993superior}
V.~Puel, M.~Caudry, P.~Le~M{\'e}tayer, J.-C. Baste, D.~Midy, C.~Marsault,
  H.~Demeaux, and J.-P. Maire.
\newblock Superior vena cava thrombosis related to catheter malposition in
  cancer chemotherapy given through implanted ports.
\newblock {\em Cancer}, 72(7):2248--2252, 1993.

\bibitem{renard2020getfem}
Y.~Renard and K.~Poulios.
\newblock {GetFEM: Automated FE modeling of multiphysics problems based on a
  generic weak form language}.
\newblock {\em ACM Transactions on Mathematical Software}, 47(1):1--31, 2020.

\bibitem{rohan2018modeling}
E.~Rohan, V.~Luke{\v{s}}, and A.~Jon{\'a}{\v{s}}ov{\'a}.
\newblock {Modeling of the contrast-enhanced perfusion test in liver based on
  the multi-compartment flow in porous media}.
\newblock {\em Journal of Mathematical Biology}, 77(2):421--454, 2018.

\bibitem{schneider2014tgif}
M.~Schneider, S.~Hirsch, B.~Weber, G.~Sz{\'e}kely, and B.~Menze.
\newblock {TGIF: Topological gap in-fill for vascular networks}.
\newblock In {\em International Conference on Medical Image Computing and
  Computer-Assisted Intervention}, pages 89--96. Springer, 2014.

\bibitem{schneider2012tissue}
M.~Schneider, J.~Reichold, B.~Weber, G.~Sz{\'e}kely, and S.~Hirsch.
\newblock {Tissue metabolism driven arterial tree generation}.
\newblock {\em Medical Image Analysis}, 16(7):1397--1414, 2012.

\bibitem{secomb2013angiogenesis}
T.~Secomb, J.~Alberding, R.~Hsu, M.~Dewhirst, and A.~Pries.
\newblock {Angiogenesis: An adaptive dynamic biological patterning problem}.
\newblock {\em PLoS Computational Biology}, 9(3):e1002983, 2013.

\bibitem{shipley2020hybrid}
R.~J. Shipley, A.~F. Smith, P.~W. Sweeney, A.~R. Pries, and T.~W. Secomb.
\newblock {A hybrid discrete--continuum approach for modelling microcirculatory
  blood flow}.
\newblock {\em Mathematical Medicine and Biology: A Journal of the IMA},
  37(1):40--57, 2020.

\bibitem{stergiopulos1996four}
N.~Stergiopulos, B.~Westerhof, J.-J. Meister, and N.~Westerhof.
\newblock {The four-element windkessel model}.
\newblock In {\em Proceedings of 18th Annual International Conference of the
  IEEE Engineering in Medicine and Biology Society}, volume~4, pages
  1715--1716. Ieee, 1996.

\bibitem{stoverud2012modeling}
K.~St{\o}verud, M.~Darcis, R.~Helmig, and M.~Hassanizadeh.
\newblock {Modeling concentration distribution and deformation during
  convection-enhanced drug delivery into brain tissue}.
\newblock {\em Transport in Porous Media}, 92(1):119--143, 2012.

\bibitem{vidotto2019hybrid}
E.~Vidotto, T.~Koch, T.~K{\"o}ppl, R.~Helmig, and B.~Wohlmuth.
\newblock {Hybrid models for simulating blood flow in microvascular networks}.
\newblock {\em Multiscale Modeling \& Simulation}, 17(3):1076--1102, 2019.

\bibitem{vignati2012fully}
A.~Vignati, V.~Giannini, A.~Bert, P.~Borrelli, M.~De~Luca, L.~Martincich,
  F.~Sardanelli, and D.~Regge.
\newblock A fully automatic multiscale 3-dimensional hessian-based algorithm
  for vessel detection in breast dce-mri.
\newblock {\em Investigative radiology}, 47(12):705--710, 2012.

\bibitem{wu2020patient}
C.~Wu, D.~A. Hormuth, T.~A. Oliver, F.~Pineda, G.~Lorenzo, et~al.
\newblock {Patient-specific characterization of breast cancer hemodynamics
  using image-guided computational fluid dynamics}.
\newblock {\em IEEE Transactions on Medical Imaging}, 39(9):2760--2771, 2020.

\bibitem{wu2019quantitative}
C.~Wu, F.~Pineda, D.~A. Hormuth, G.~S. Karczmar, and T.~E. Yankeelov.
\newblock {Quantitative analysis of vascular properties derived from ultrafast
  DCE-MRI to discriminate malignant and benign breast tumors}.
\newblock {\em Magnetic Resonance in Medicine}, 81(3):2147--2160, 2019.

\end{thebibliography}
